\documentclass[aps,amsmath, twocolumn]{revtex4}
\usepackage[dvips]{graphicx}
\usepackage{epsfig}
\usepackage{dcolumn}
\usepackage{bm}
\usepackage[dvips]{color}

\begin{document}

\title{Intrinsic and collective effects in assemblies of nanomagnets}
\author{H. Kachkachi}
\email{hamid.kachkachi@physique.uvsq.fr}
\homepage{http://hamid.kachkachi.free.fr}
\affiliation{Groupe d'Etude de la Mati\`ere Condens\'ee, Universit\'e de
Versailles St. Quentin, 45 av. des Etats-Unis, 78035 Versailles, France}

\begin{abstract}
In this paper we review some results of our works on the magnetization processes in: i) Isolated nanomagnets, both in the one-spin approximation and as many-spin systems. Here, we focus on the intrinsic properties, e.g., those induced by finite-size, boundary and surface effects. We also investigate the crossover between the two regimes. ii) Assemblies of nanomagnets, also in the two situations. We focus on their behavior mainly due to dipole-dipole interactions. Then, we will comment on the interplay between these intrinsic and collective effects.
\end{abstract}

\date{\today}
\maketitle
\tableofcontents

\section{\label{intro}Introduction}
Magnetic information storage is one of the major applications of magnetic nanoparticles, though magnetic nanoscale structures in general have been identified as offering a great potential in applications ranging from nano “spin” electronics, innovative complex materials like magnetically controlled shape memory alloys to medical applications in medical diagnosis and therapy. The objective of achieving high-density requires rather small sizes, high anisotropies, and imposes severe constraints on the microstructure and the magnetic properties to ensure a good stability against thermal demagnetization over several years.
On the other hand, the same material must be magnetized by the writing head in a time interval as short as a nanosecond. However, smaller sizes induce stronger surface effects and higher densities induce stronger inter-particle interactions. Unfortunately, surface effects and inter-particle interactions have the drawback to negatively affect the signal-to-noise ratio. Optimising the underlying material in view of technological applications requires taking account of and understanding the effects of surface and interactions, both on the static and dynamic level.

At present there are mainly two types of nanoparticle samples: i) Assemblies of nanoparticles embedded in a non-magnetic matrix with volume distribution and randomly oriented easy axes, with negligible to strong dipole-dipole inter-particle interactions (DDI) \cite{doretal97acp}.
ii) Isolated single particles measured by the technique of $\mu$-SQUID \cite{wernsdorfer01acp}.
From a theoretical point of view, none of these sample prototypes is  so far fully satisfactory: While in the case i) one can measure by commercial SQUID most of the magnetic properties, e.g., field-cooled (FC), zero-field-cooled (ZFC) magnetization and ac susceptibility, in a wide temperature range (4.2-300K) and in any applied magnetic field available (0-30T), this case presents tremendous difficulties for modeling because of the volume and anisotropy easy axes distributions and DDI.
Indeed, these interactions, because of their long-range character, together with the two distributions, require prohibitively long CPU times to deal with realistic assemblies of, say $10^4$ particles of $300-1000$ spins each. Besides, the physical quantities, such as the magnetization of the assembly, are only known on average.
In the case ii), one gets rid of these difficulties but then the experimental results are limited to low temperature and applied field, and more importantly the magnetization itself is not accessible, and for very small particles the signal is rather ``noisy", which hinders any measurement of hysteresis loops, and one makes do with the angular dependence of the switching field, the so-called {\it Stoner-Wohlfarth astroid}.

In many theoretical approaches the magnetic state of a small magnetic particle is represented by a single magnetic moment regardless of its microscopic origin, this is the {\em one-spin approximation}, or one-spin problem (OSP), which can only be justified for particles that are not too large to be single-domain, and not too small to be surface-effects free.
Within this approximation, it is well known by now that the magnetization of a nanoparticle can overcome the anisotropy-energy barrier and thus reverse its direction, at least in two ways: either under applied magnetic field which suppresses the energy barrier, or through thermally activated statistical fluctuations.
Switching under applied field, at zero temperature, is described by the Stoner-Wohlfarth (SW) model \cite{stowoh48, stowoh91}.
In fact, it has recently been shown \cite{shuetal03prl1, shuetal03prl2}, experimentally and theoretically, that efficient magnetization switching can also be triggered by transverse field pulses of a duration that is half the precession period.
The SW model accounts for the hysteretic reversal of a magnetic moment over the potential barrier under the influence of the magnetic field applied in an arbitrary direction and at zero temperature.
In fact, at zero temperature the magnetization switching is only possible when the field reaches a critical value that  suppresses the energy barrier.
At finite temperature and short-time scales, crossing of the energy barrier is triggered by statistical thermal fluctuations and is described by the N\'eel-Brown (NB) model \cite{nee49ag, nee49cras, bro63pr, bro79ieee} and its extensions, which took several years to be fully accomplished [see the review article \cite{coffeyetal01acp} and references therein].
Both the SW and NB models have been confirmed, to some extent, by experiments on individual cobalt particles \cite{wernsdorfer01acp} through measurements of the switching field distribution, the non-switching probability, and the time evolution of the direction of the particle's magnetization.
At finite temperature, but at quasi-equilibrium, the magnetization switching occurs according to two distinct regimes \cite{kacgar01physa291}: At very low temperature, switching operates by coherent rotation of all spins, as in the SW model, whereas at higher temperatures, the magnetization switches by changing its magnitude.
The latter effect results in a shrinking of the SW astroid as qualitatively described by the modified Landau theory \cite{kacgar01physa291}, and confirmed by experiments \cite{wernsdorfer01acp}.

However, it is clear that the change of the magnetization magnitude during its switching cannot be explained in the framework of the OSP approximation. Indeed, it can only be understood as the result of a successive switching of individual (or clusters of) spins inside the particle \cite{kacdim02prb, dimkac02jap}, which should then be necessarily dealt with as a many-spin system.
In more general terms, the picture of a single-domain magnetic particle with all spins pointing into the same direction, leading to coherent relaxation processes becomes unsatisfactory as soon as one comes to deal with small particles with a surface that makes up to $50\%$ of the total volume. This is to say, the surface entails strong effects that cannot be neglected.
In fact, deviations from the OSP approximation, and thereby from both the SW and NB models, have been observed in metallic particles \cite{cheetal95prb, respaudetal98prb}, ferrite particles \cite{ricetal91jap, kodber99prb}, and maghemite particles~\cite{troetal00jmmm}.
These deviations have materialized in terms of the absence of magnetization saturation at high fields, shifted hysteresis loops after cooling in field, and enhancement at low temperature of the magnetization as a function of the applied field.

In most cases, surface effects happen to be strong enough as to compensate for the work needed against the exchange energy that favors full alignment, and it is conceivable then that the magnetization vector points along the easy axis in the core of the particle, and then gradually turns into a non-collinear direction when it approaches the surface.
Moreover, in addition to exchange interactions, which are the strongest interactions between atomic moments in a magnetic system, there are also the purely magnetic dipole interactions between the magnetic moments of the atoms and the interactions between the magnetic moments and the electric field of the crystal lattice (spin-orbit interactions).
The last two types of interactions, being relativistic in origin, are much weaker than the exchange energy on a short-distance scale. However, on a long-distance scale they are non-negligible because they are of long range.
These interactions have two important features: they induce inhomogeneities in the spatial distribution of the magnetization, and introduce a preferred direction in the system \cite{akhbarpel68}.
Roughly, DDI lead to two energy terms corresponding to the volume and surface charges. In a small
magnetic system, such as a nanoparticle, only the second contribution is significant and it accounts for the shape anisotropy, and the former becomes negligible as one integrates over a small volume \cite{hah70prb, aha96}.
Therefore, to study a small magnetic system where inhomogeneities are necessarily present, one has to take into account all different contributions to the energy: exchange and DDI, core and surface anisotropies, and Zeeman energy.

So, one of our goals is to understand surface effects on the thermodynamic and spatial behavior of the magnetization in small systems. This is also of crucial importance to the study of their dynamics.
However, this requires a microscopic approach that accounts for the local environment inside the particle and the above mentioned contributions.
Unfortunately, this leads to a rather difficult, if not insuperable, task owing to the large number of degrees of freedom which hinders any attempt to analyze the energyscape whose knowledge is indispensable for understanding the dynamics and switching mechanisms.
For this reason, inter alia, calculations of the reversal time of the magnetization of fine single-domain ferromagnetic particles, for instance, initiated by N\'eel \cite{nee49ag, nee49cras}, and set firmly in the context of the theory of stochastic processes by Brown \cite{bro63pr, bro79ieee, bro63mic}, have invariably proceeded by ignoring all kind of interactions. Thus, the only terms which
are taken into account in these calculations are the internal magneto-crystalline anisotropy of the particle, the random field due to thermal fluctuations, and the Zeeman term \cite{coffeyetal01acp}.
However, before attacking the general problem of the effect of interactions and that of the surface on the dynamical properties, one has to gain a sufficient understanding of the static properties.

Yet before tackling the effect of surface anisotropy, one has to clearly define, understand, and when possible evaluate in a separate way, finite-size effects (artificially) induced in theoretical calculations by considering a
system of finite size, and ``true" boundary effects (naturally) induced by crystal field symmetry breaking on the surface.
Finite-size effects in magnetic systems have been of much interest for decades now, and have been extensively studied by many authors~\cite{fisbar72prl, fispri85prb, fispri86cmp, barfis73ap, bin92fss, binher92}, just to cite a few.
One of the difficulties, inherent to systems of ``round" (spherical or ellipsoidal) geometries, as is the case with nanoparticles, is the separation of boundary effects due to defects in coordination on the boundary and the unavoidable finite-size effects.
In hypercubic systems, this problem is handled by using periodic and free boundary conditions, but this is not possible in other topologies, and thus boundary and finite-size effects are tangled together.
More difficulties come into play when the spins on the surface are attributed some anisotropy.
While for thin films and layered systems the direction of the anisotropy on the surface is clearly determined and its intensity estimated~\cite{bruren89apa, frietal94jmmm, stoeffler97jmmm}, especially in  $3d$ elements, for various thicknesses, no definite answer is yet available for nanoparticles.
Nevertheless, there are some estimations of the surface anisotropy constant~\cite{shilov99phd, doretal96prb, gazetal97epl, kodetal96prl} obtained in an indirect way by fitting magnetic measurements and assuming an effective anisotropy constant.
Unfortunately, today the experimental techniques are not so advanced as to allow for probing the crystallographic structure of a nanoparticle surface, in the same manner as is done for $2d$ magnetism, and hence no information is yet available as to the precise nature of the anisotropy on the surface.
As such, theorists have no definite answer as to what is the most adequate model for describing surface anisotropy in nanoparticles.
To the best of our knowledge there are mainly two models with positive or negative constants: The {\it transverse surface anisotropy} (TSA)~\cite{bro63mic} model where a surface spin is attributed a single-site anisotropy axis that is normal to the surface, and N\'eel's model (NSA)~\cite{nee53cras, nee54jpr, vicmac93prb, chubaloha94prb}.
One may also find in the literature a model with random anisotropy.
N\'eel's model is more physically plausible because the anisotropy at a site occurs only if the latter presents defects in its environment, e.g., if it lacks some of its neighbors.

When a nanoparticle is embedded in an assembly dispersed in a non-magnetic matrix (of polymer, silica, etc.), still more interesting challenging physical phenomena are observed, such as the disappearance of the maximum of the temperature at the peak of the zero-field-cooled magnetization of an interacting assembly, the magnetization enhancement at low temperature and high field in dilute maghemite particles~\cite{ezzir98phd, troetal00jmmm}, and the appearance of spin-glass states at low temperature in concentrated assemblies~\cite{troetal00jmmm}.
In addition to the volume and anisotropy easy axes distributions, particle assemblies also introduce a new physical parameter, namely the inter-particle interactions, as discussed earlier, and in particular DDI. These interactions have been studied for many decades in different areas of physics, and in particular in magnetism~\cite{holpri40pr}.
There is an overwhelming literature investigating the influence of DDI on both static and dynamic properties of nanoparticles, and it would be almost impossible to provide a fair account of all of them.
However, we do not know of theoretical work that deals with DDI between particles taking at the same time account of their internal structure and thereby surface effects.
In Ref.~\onlinecite{jongar01prb}, the Landau-Lifshitz thermodynamic perturbation theory~\cite{lanlif80} is used to tackle the case of weakly dipolar-interacting monodisperse assemblies of magnetic moments, in the OSP approximation, with uniformly or randomly distributed anisotropy axes. The authors studied the influence of DDI on the susceptibility and specific heat of the assembly.
On the other hand, Monte Carlo technique has been used to study the effect of DDI (of arbitrary intensity) on the temperature behavior of the coercive field and remanent magnetization in a uniform assembly of magnetic moments~\cite{kectro98prb}.
The situation involving both surface effects and DDI has never been considered so far because of its tremendous complexities. Needless to say that, already at the static level, no exact analytical treatment of any kind is ever possible even in the OSP approximation, i.e., ignoring the internal structure of the particles.

This review is organized as follows: In section II, which deals with the intrinsic features of nanoparticles, we emphasize the difference between finite-size, boundary and surface effects. Then we introduce our many-spin approach and discuss and compare two surface anisotropy models (Transverse and N\'eel). Next, we present and interpret some of the main results on the static behavior of many-spin particles, and in particular on the role of surface anisotropy. We end this section with a discussion of the crossover between the OSP and MSP regimes.
Section II deals with the collective behavior of nanoparticle assemblies with special emphasis on DDI: i) we summarize our work on the effect of anisotropy and weak DDI on the magnetization of a polydisperse assembly of magnetic moments, i.e., OSP particles. We provide the corresponding approximate analytical expressions and compare them with Monte Carlo simulations. ii) We investigate the effect of DDI on the ZFC magnetization, and in particular on the temperature at its maximum as a function of the applied magnetic field. iii) We end this section with a very short discussion, together with a preliminary result, of assemblies of non-interacting MSP  particles.
\section{\label{Intrinsic}Intrinsic effects: surface anisotropy}
As discussed in the introduction, when dealing with fine magnetic systems, such as a nanomagnet, one should distinguish, at least from a theoretical point of view, between finite-size, boundary, and surface effects.

\subsection{Finite-size versus boundary effects}
\begin{figure*}[floatfix]
\includegraphics[width=5cm]{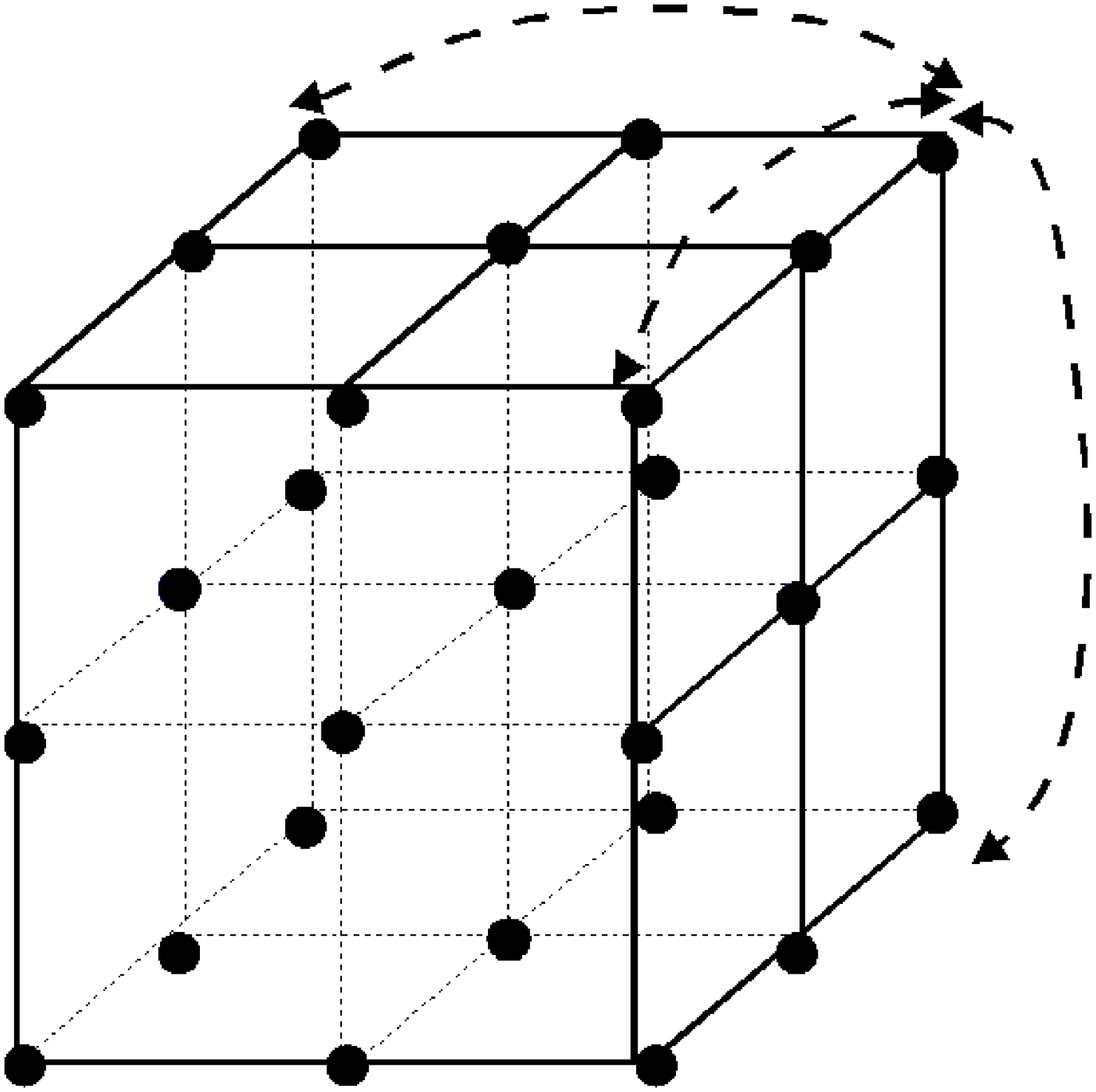} \hspace{0.5cm}
\includegraphics[width=7cm]{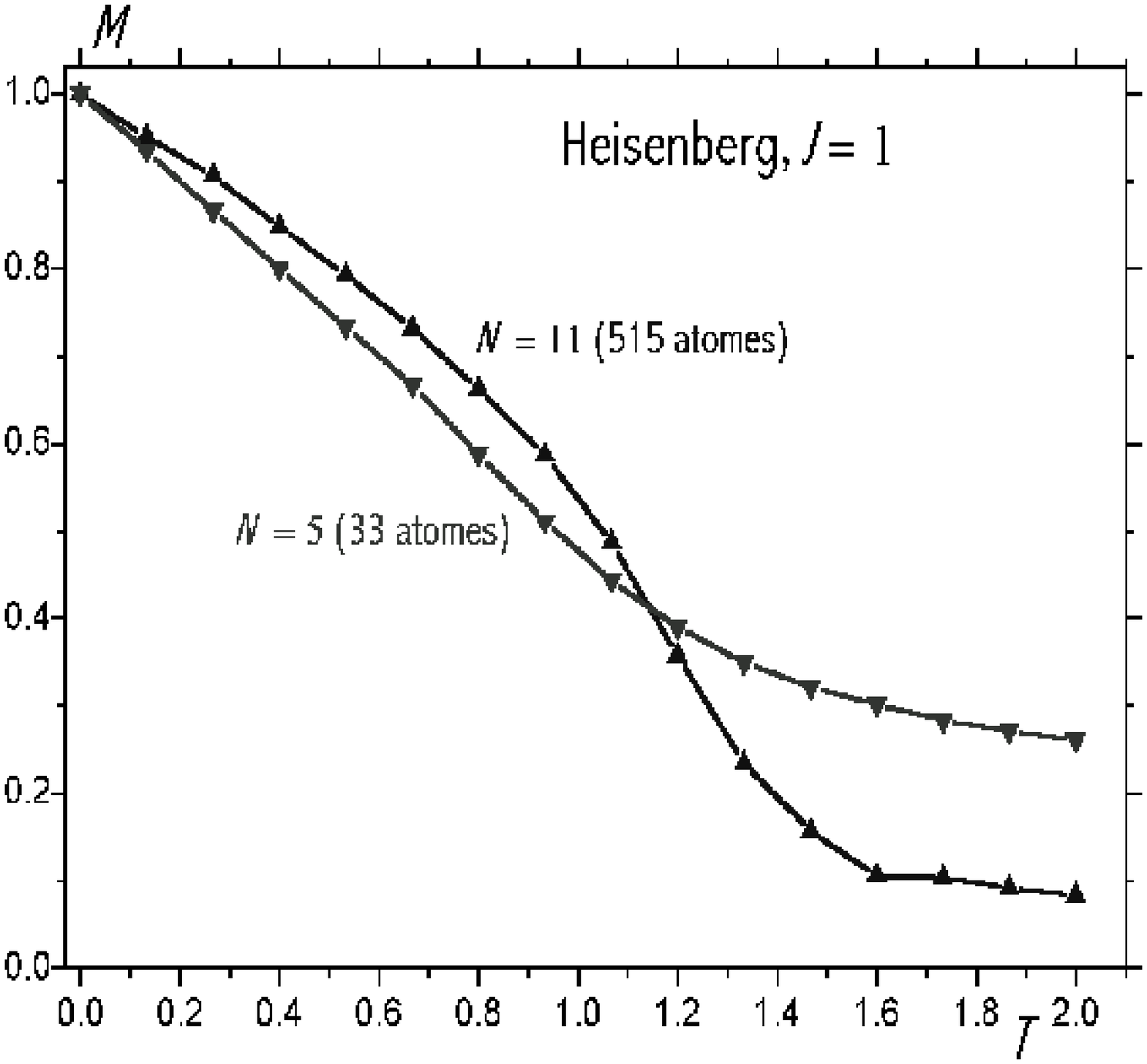}
\caption{Cubic system with pbc and thermal behavior of the magnetization for
two particle sizes.}
\label{pbc}
\end{figure*}

For instance, for a simple cubic (sc) lattice (see Fig.~\ref{pbc} left) with periodic boundary conditions (pbc), there is only one environment (crystal field) with coordination number $z=6$. 
%
%
In this case, the temperature behavior of the magnetization is marked by the well-known $M\sim 1/\sqrt{\mathcal{N}}$ tail in the critical region, where $\mathcal{N}$ is the total number of spins in the particle (Fig.~\ref{pbc}, right). 
%
\begin{figure*}[floatfix]
\includegraphics[width=5cm]{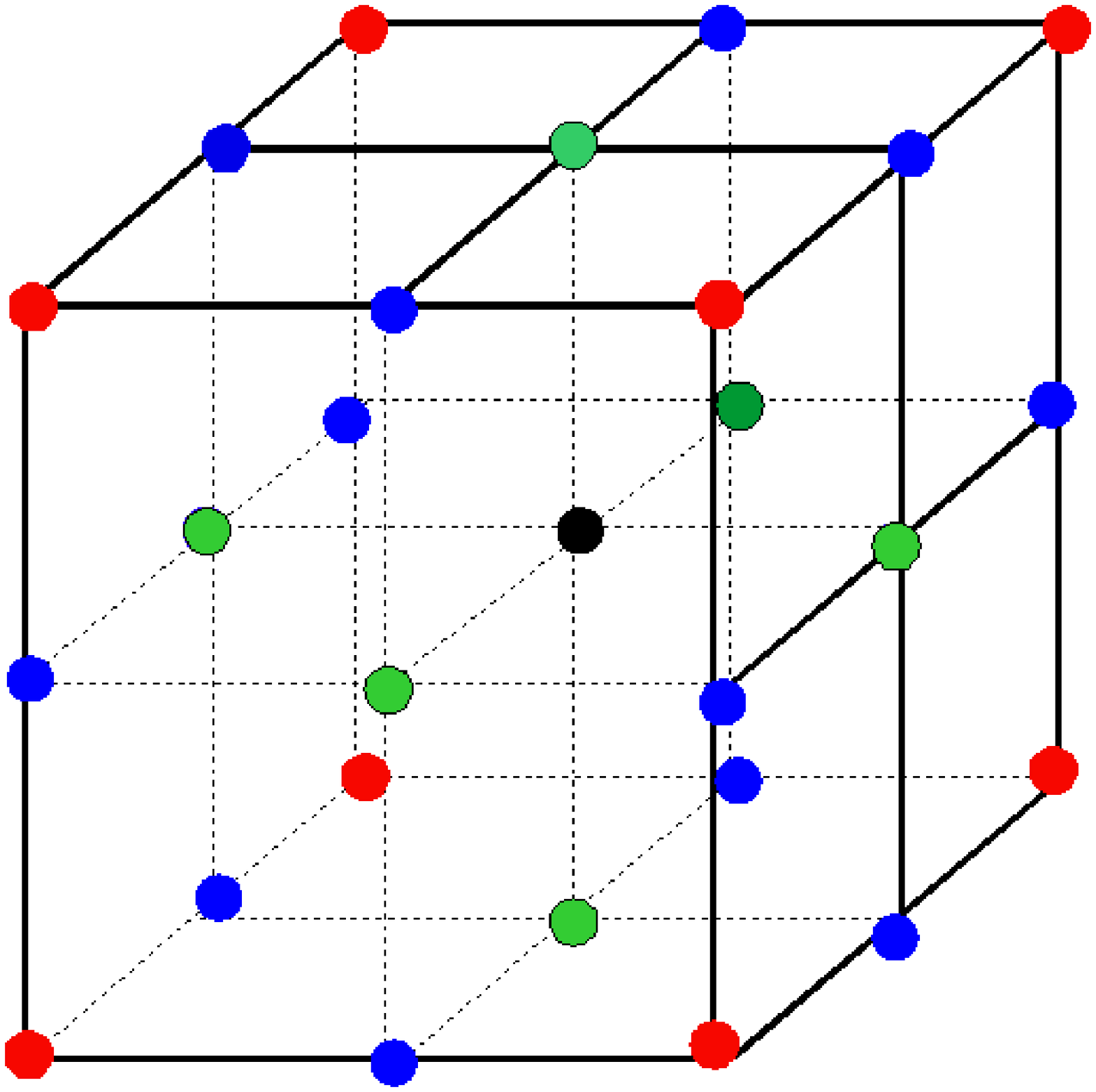} \hspace{0.5cm}
\includegraphics[width=8cm]{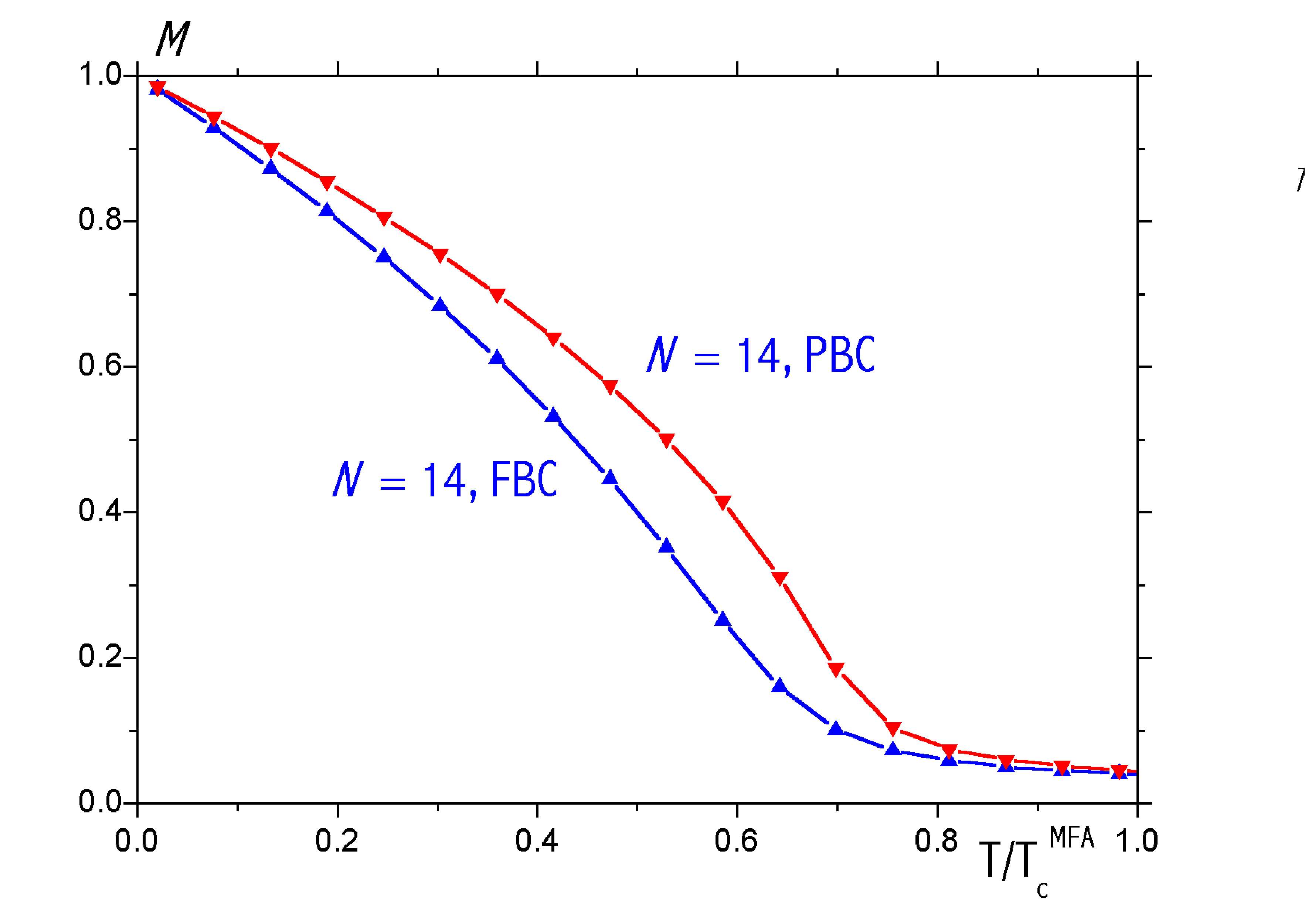}
\caption{Cubic system with fbc and thermal behavior of the magnetization for
pbc and fbc systems.}
\label{fbc}
\end{figure*}
%
In the case of more realistic free boundary conditions (fbc), a cube with sc structure shows four different environments with $z=3,4,5,6$ (see Fig.~\ref{fbc} left). In this case, in addition to the previous finite-size effects, one is faced with boundary effects. These induce stronger fluctuations that suppress the magnetization of the system (see Fig.~\ref{fbc} right). 
Considering both cases of pbc and fbc allows for a separate estimation of the related effects. Now, if the boundary of a system with fbc is endowed with a surface anisotropy, which is indeed a consequence of boundary defects, we may then speak of surface effects, in addition to the finite-size and boundary effects.

For both pbc and fbc, we have shown \cite{kacgar01physa300} that the magnetization can be written in a simple form. At low temperatures and zero field $M$ deviates from 1, its saturation value, according to 
\begin{equation}  \label{MLowTH0}
M\cong 1-\frac{\theta}{2} W_{N},
\end{equation}
where 
\begin{equation}  \label{WN}
W_{N}=\frac{1}{\mathcal{N}}\sum_{\mathbf{k}}{}^{^{\prime }}\frac{1} {%
1-\lambda _{\mathbf{k}}},
\end{equation}
and for a three-dimensional ($d=3$) sc lattice $\lambda_{\mathbf{k}}=(\cos k_{x}+\cos k_{y}+\cos k_{z})/d$. It is important to note that $W_{N}$ in (\ref{WN}) for pbc and fbc differ only by the definition of the discrete wave vectors, since \cite{kacgar01physa300, kacgar01epjb} 
\begin{equation}  \label{defkpbc}
k_{\alpha }=\left\{ 
\begin{array}{cc}
2\pi n_{\alpha }/N, & \mathrm{pbc} \\ 
\pi n_{\alpha }/N, & \mathrm{fbc}
\end{array}
\right. ,\qquad n_{\alpha }=0,1,...,N-1
\end{equation}
where $\alpha =x,y,z.$ This subtle difference is responsible for much stronger thermal fluctuations in the fbc model due to boundary effects. The difference between the sums $W_{N}$ (with finite $N$, which is the side of the cube) and the so-called Watson's integral (for bulk)
\begin{equation}  \label{Watson} 
W=\int \!\!\!\frac{d^{3}\mathbf{k}}{(2\pi )^{3}}\frac{1}{1-\lambda _{\mathbf{k}}},
\end{equation}
with $W=1.51639$ for $d=3$, describes the finite-size effects for the pbc case \cite{fispri85prb, fispri86cmp} and boundary effects in the fbc case  \cite{kacgar01physa300}. Indeed, we have \cite{kacgar01physa300}
\begin{equation}  \label{DeltaN}
\Delta _{N}\equiv \frac{W_{N}-W}{W}\cong \left\{
\begin{array}{ll}
\displaystyle -\frac{0.90}{N}, & \mathrm{pbc} \\ 
&  \\ 
\displaystyle \frac{9\ln (1.17N)}{2\pi WN}, & \mathrm{fbc}
\end{array}
\right.
\end{equation}
%
\begin{figure}[floatfix]
\includegraphics[angle=-90,width=8cm]{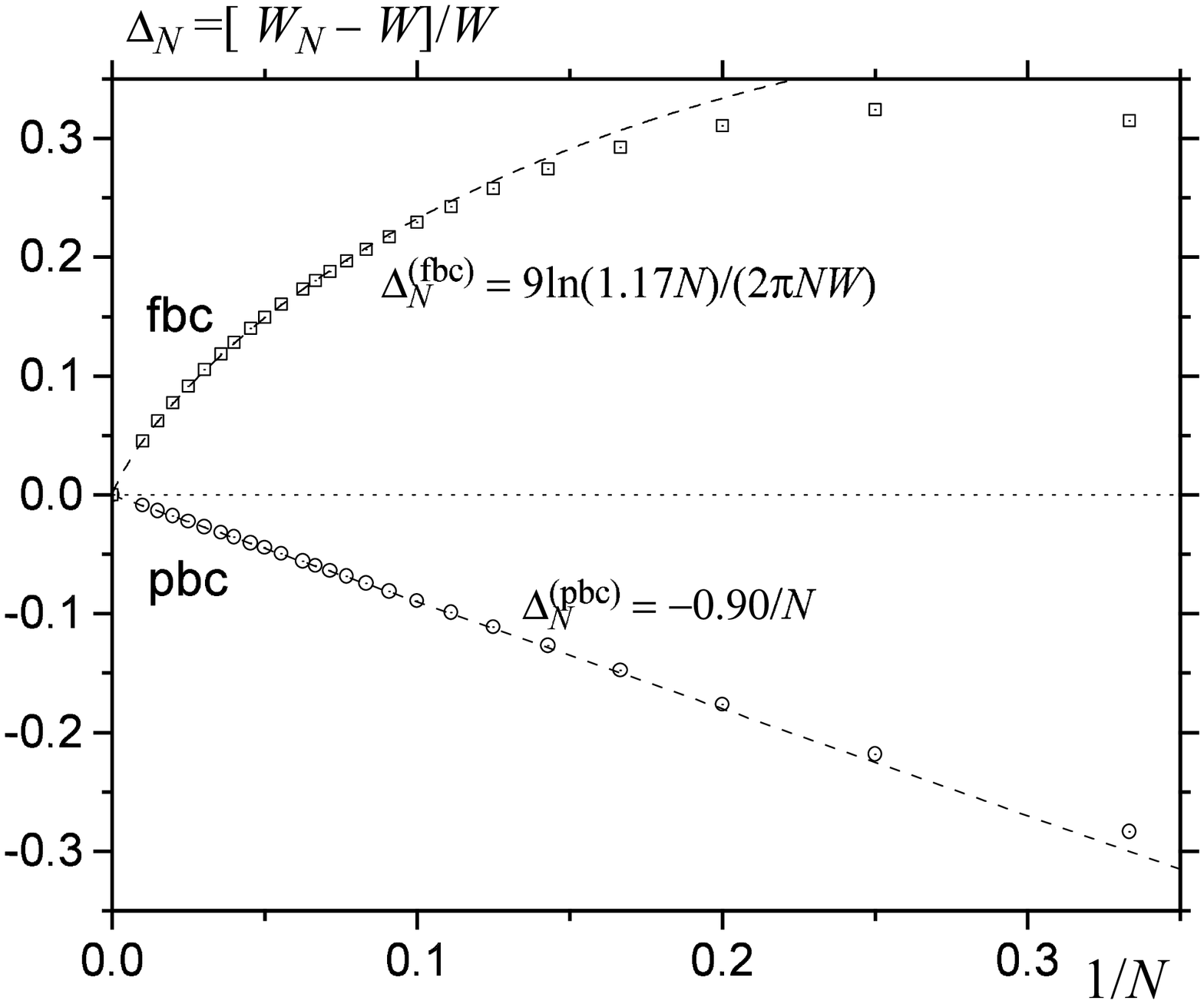}
\caption{Lattice sums $W_N$ for cubic systems with free and periodic
boundary conditions. $W=1.51639$ is the bulk value for the sc lattice.}
\label{sef_del}
\end{figure}

Therefore, Fig.~\ref{sef_del} shows that the coefficient in the linear-$\theta $ term in Eq.~(\ref{MLowTH0}) is smaller than in the bulk for the pbc system and greater for the fbc system. Consequently, this means that the
boundary effects suppress the magnetization while finite-size effects enhance it. 
Moreover, Fig.~\ref{sef_del} also shows that boundary effects render a larger contribution than finite-size effects, making the net magnetization below that of the bulk. 
%
%
\subsection{\label{ESA}Surface effects : MSP approach}
Surface effects are due to the breaking of crystal-field symmetry at the boundary of the nanoparticle, and in reality they cannot be separated from boundary effects. In order to study
such effects one has to resort to microscopic theories, unlike the macroscopic SW model \cite{stowoh48, stowoh91}, capable of distinguishing
between different atomic environments and taking account of physical parameters such as single-site surface anisotropy, exchange and DDI, in addition of course to the magneto-crystalline anisotropy in the core and magnetic field.
Unfortunately, this leads to difficult many-body problems which can only be efficiently dealt with, in general, using numerical approaches.

In this section we will summarize some of the main results obtained so far.

\subsubsection{Model and notation}
We consider a ferromagnetic particle of $\mathcal{N}$ spins cut from a cube of
side $N$ (i.e., $N-1$ atomic spacings) with a given lattice structure.
Due to the underlying (discrete) lattice structure, the particle thus obtained is not a sphere with smooth boundary because its outer shell presents apices, steps, and facets, resulting in many sites with different coordination numbers.

The model Hamiltonian we use is the (classical) anisotropic Dirac-Heisenberg model~\cite{kacetal00epjb, kacdim02prb}
\begin{equation}
\mathcal{H} = -\sum\limits_{\left\langle i,j\right\rangle }J_{ij}\mathbf{s}_{i}\cdot
\mathbf{s}_{j}-(g\mu _{B})\mathbf{H}\cdot \sum\limits_{i=1}^{\mathcal{N}}\mathbf{s}
_{i} + \mathcal{H}_{an},
\end{equation}
where $\mathbf{s}_{i}$ is the unit spin vector on site $i$, $\mathbf{H}$ the
uniform magnetic field, $\mathcal{N}$ the total number of spins (core and surface),
and $J_{ij}(=J>0)$ the nearest-neighbor ferromagnetic exchange coupling. $\mathcal{H}_{an}$ is the uniaxial single-site anisotropy energy
\begin{equation}  \label{uaa}
\mathcal{H}_{an}=-\sum\limits_{i}K_{i}(\mathbf{s}_{i}\cdot \mathbf{e}_{i})^{2},
\end{equation}
with easy axis $\mathbf{e}_{i}$ and constant $K_{i}>0$. If the spin at site $i$ is in the core, the anisotropy axis $\mathbf{e}_{i}$ is taken along the reference $z$ axis and $K_{i}=K_{c}$. For surface spins, this axis is along the radial (i.e., transverse to the cluster surface) direction and $K_{i}=K_{s}$. In this case, the model in (\ref{uaa}) for surface spins is called the TSA model introduced earlier.
We use the more general model of transverse direction given by the gradient [the vector perpendicular to the isotimic surface $\psi=\mathrm{constant}$ defining the shape of the particle, e.g. a sphere or an ellipsoid]. In the case of a spherical particle the transverse and radial directions coincide, whereas for another geometry such as an ellipsoid they do not.

A more physically appealing microscopic model of surface anisotropy was
introduced by N\'eel \cite{nee53cras} with
\begin{equation}\label{NSA}
\mathcal{H}_{an}^\text{N\'eel} = \dfrac{K_s}{2}\sum\limits_{i}\sum\limits_{j=1}^{z_i}({\bf s}_{i}\cdot{\bf u}_{ij})^{2},
\end{equation}
where $z_i$ is the coordination number of site $i$ and ${\bf u}_{ij}={\bf r}_{ij}/r_{ij}$ is the unit vector connecting the site $i$ to its nearest
neighbors $j=1,\ldots,z_i$. This model is more realistic since the anisotropy at a given site occurs only when the latter loses some of its neighbors, i.e., when it is located on the boundary.
The model in (\ref{NSA}) is referred to as the N\'eel surface anisotropy (NSA) model \cite{garkac03prl}.

Qualitatively, the NSA model is not quite different from the TSA model. 
For example, consider a site $i$ sitting on a $[100]$ facet, e.g. in the upper most plane normal to the $z$ axis. It has $4$ neighbors on that facet and one below it along the $z$ axis. 
From (\ref{NSA}), the corresponding energy reads \cite{kacbon06prb}
\begin{widetext}
\begin{eqnarray*}
\mathcal{H}_i^\mathrm{NSA} &=& K_s\left[ \left( \mathbf{s}_{i}\cdot \mathbf{e
}_{x}\right) ^{2} + \left( \mathbf{s}_{i}\cdot -\mathbf{e}_{x}\right)
^{2} + \left( \mathbf{s}_{i}\cdot \mathbf{e}_{y}\right)^{2}+\left( \mathbf{s}
_{i}\cdot-\mathbf{e}_{y}\right)^{2}
+ \left( \mathbf{s}_{i}\cdot -\mathbf{e}_{z}\right)^{2}\right] \\
&=& K_s\left[2s_{i,x}^{2} + 2s_{i,y}^{2} + s_{i,z}^{2}\right] 
= 2 K_s - K_s\,s_{i,z}^{2},
\end{eqnarray*}
\end{widetext}
where we have used $\parallel{\bf s}_i\parallel=1$. This implies that if $K_s>0$ the easy direction is along $\pm \mathbf{e}_{z}$, i.e., normal to the facet, and if $K_s<0$ the facet becomes an easy plane. 
Therefore, upon dropping the irrelevant constant, we rewrite the above energy as
\[
\mathcal{H}_i^\mathrm{NSA} = -K_s\,\left(\mathbf{s}_{i}\cdot \mathbf{e}_{z}\right)^{2}. 
\]
which is the same as the TSA in Eq.~(\ref{uaa}) for the site considered. 
More generally, averaging the NSA over a surface perpendicular to the direction $\mathbf{n}$ leads to [see Eqs.~(6, 7) in Ref.~\onlinecite{garkac03prl}]
\begin{equation}
\mathcal{H}_i^\mathrm{NSA} = - K_s\left(|n_x|s_x^2 + |n_y|s_y^2 + |n_z|s_z^2\right),
\end{equation}
thus favoring among $(x, y, z)$ the direction closest to the surface normal. This explains the similarity between the results obtained with TSA and NSA.
Using the components $|n_\alpha|,\alpha=x,y,z$, the atomic surface density reads \cite{garkac03prl}
\begin{equation}\label{fnDef}
f(\mathbf{n})=\max\left\{|n_{x}|,|n_{y}|,|n_{z}|\right\},  
\end{equation}
and thereby we have the particular cases
\begin{equation*}
\mathcal{H}_i^\mathrm{NSA} = 
\left\{
\begin{array}{ll}
-K_s\,s_{z}^{2}, & n_{z}=1 \\
-K_s(s_{x}^{2} + s_{y}^{2})/\sqrt{2}, & n_{x} = n_{y} = \frac{1}{\sqrt{2}} \\
-K_s/\sqrt{3}, & n_{x} = n_{y} = n_{z} = \frac{1}{\sqrt{3}}.
\end{array}
\right.
\end{equation*}
For comparison, taking account of the atomic surface density, the TSA model is described by
\begin{equation}\label{ESPerp}
\mathcal{H}_i^\mathrm{TSA} = - K_s\,(\mathbf{n}\cdot \mathbf{s}_i)^{2}\,f(\mathbf{n})  
\end{equation}
and the whole effect comes from the atomic surface density. 
Quantitatively, in the NSA model the effect is bigger, since for the surface cut perpendicular to the grand diagonal of the cubic lattice the anisotropy completely disappears, whereas in (\ref{ESPerp}) the surface anisotropy is only reduced by the factor $1/\sqrt{3}$.
\subsubsection{Results and discussion}
The static properties of magnetic nanoparticles as many-spin systems have been studied in Refs.~\cite{kacetal00epjb, kacgar01physa300, kacgar01epjb, dimkac02jap, kacdim02prb, garkac03prl, kacmah04jmmm} using the classical Heisenberg model involving the atomic spin, as the elementary building block, with continuous degrees of freedom.
In these works we investigated the field, temperature, and spatial behavior of the net as well as the local magnetization.
We adapted and applied the existing theoretical approaches, such as spin-wave theory, spherical model, Monte Carlo simulation, and numerical techniques to nano-scaled magnetic systems.
In particular, we studied the magnetic structure and found new switching mechanisms of the particle endowed with surface anisotropy.
It was possible to estimate and compare finite-size, boundary, and surface effects, and to show that the latter two effects are of long-range and progressive, and that surface effects induce non-saturation of the magnetization, as observed in experiments  \cite{troetal00jmmm, respaudetal98prb, cheetal95prb}.
In the particular (and typical) situation where the exchange interaction is much stronger than the anisotropy energy we found that the magnetization can be represented as the spatially homogeneous ``global" magnetization plus a small inhomogeneous contribution that we calculated analytically and numerically \cite{garkac03prl, kacbon06prb}. 
The latter is induced by surface anisotropy and it is maximal near the surface but can extend deeply into the body of the particle. It describes the adjustment of the magnetization to the conditions at the surface by minimizing the total energy with fixed direction of the global magnetization. As a result we obtained the effective particle's energy that depends on the orientation of its net magnetization and arises because of surface anisotropy. This contribution is of second order in the surface anisotropy and it adds to other terms, such as the bulk anisotropy and the first-order contribution from the surface anisotropy, which disappear in samples of cubic or spherical shape.
These contributions to the energy of a magnetic nanoparticle are crucial to its dynamical behavior, in particular, in the ferromagnetic resonance (FMR). Accurately taking all of them into account should make it possible to determine the bulk and surface anisotropies from the experimental data. An interesting problem is the dynamical aspect of the magnetization adjustment mentioned above. If the anisotropies are much smaller than exchange interaction, the exchange-driven adjustment is much faster than the global precession of the magnetization induced by the anisotropy. Then, these adjustment modes behave adiabatically at low frequencies and the effective OSP energy is a good approximation.
For materials with a very strong surface anisotropy such separation of dynamical scales is no longer valid, and the dynamics of such nanoparticles becomes an essentially many-body process. FMR experiments on magnetic nanoparticles should allow to estimate the values of the surface anisotropy and detect different regimes of their dynamical behavior.

\begin{figure}[floatfix]
\includegraphics[angle=-90,width=12cm]{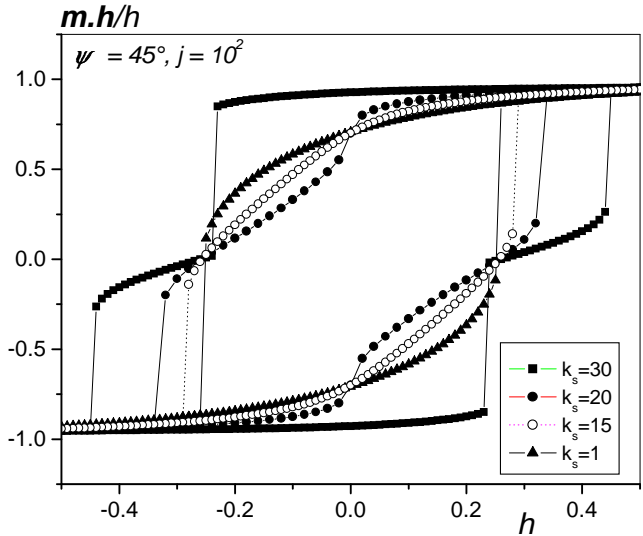}
\caption{\label{hyst_ks_a45} Hysteresis for $\psi = \pi/4,  {\cal N}=360$ and different values of surface anisotropy constant $k_{s}\equiv K_s/K_c, h=H/2K_c$.
}
\end{figure}

Let us now review with data plots a few typical results obtained for an MSP particle with core and surface anisotropies, together with intra-exchange interactions and magnetic field. 
At zero temperature, in Refs.~\onlinecite{kacdim02prb, dimkac02jap, kacmah04jmmm} we studied the magnetic state and switching mechanisms of model particles of spherical shape, sc lattice, uniaxial anisotropy in the core and both TSA and NSA, by varying the size, the exchange couplings, surface anisotropy intensity, and the field direction.
The main result of this work is that varying the surface anisotropy one observes a crossover from i) the coherent-reversal regime, as obtained by representing the particle as a macrospin according to the models of SW (for statics) and NB (for dynamics), into ii) the incoherent-reversal regime with cluster-wise switching. In the latter regime, the particle exhibits (due to strong surface anisotropy) new features that are reminiscent of a many-spin system and which cannot be described by a macroscopic approach.
Fig.~\ref{hyst_ks_a45} shows hysteresis loops for a spherical particle of ${\cal N}=360$ spins with uniaxial anisotropy (with constant $K_c$) in the core and TSA with constant $k_{s}\equiv K_s/J$. The field is applied at an angle $\psi = \pi/4$ with respect to the core easy axis.
%
%
These results show that as the surface anisotropy constant increases, one starts to observe steps in the hysteresis loop corresponding to the successive switching of spin clusters, defined as the groups of spins whose anisotropy axes make the same angle with the field direction.
Such subtle (intrinsic) features could be checked by direct measurements of the hysteresis loop of an individual nanoparticle, were they to become possible. At present, available measurements of nanoparticle assemblies do not allow for such a check. 

\begin{figure*}[floatfix]
\includegraphics[angle=-90,width=8cm]{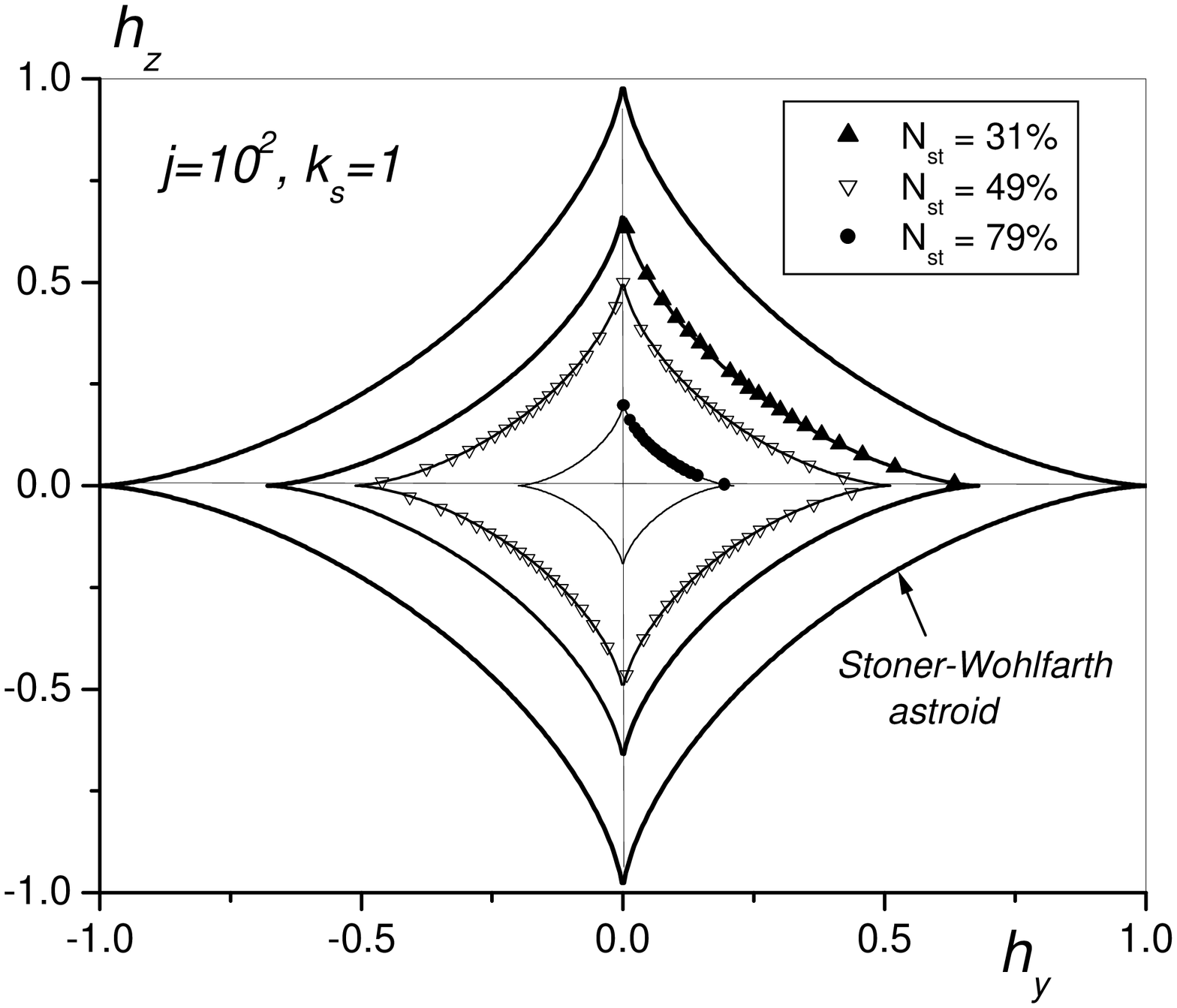}
\includegraphics[angle=-90,width=8cm]{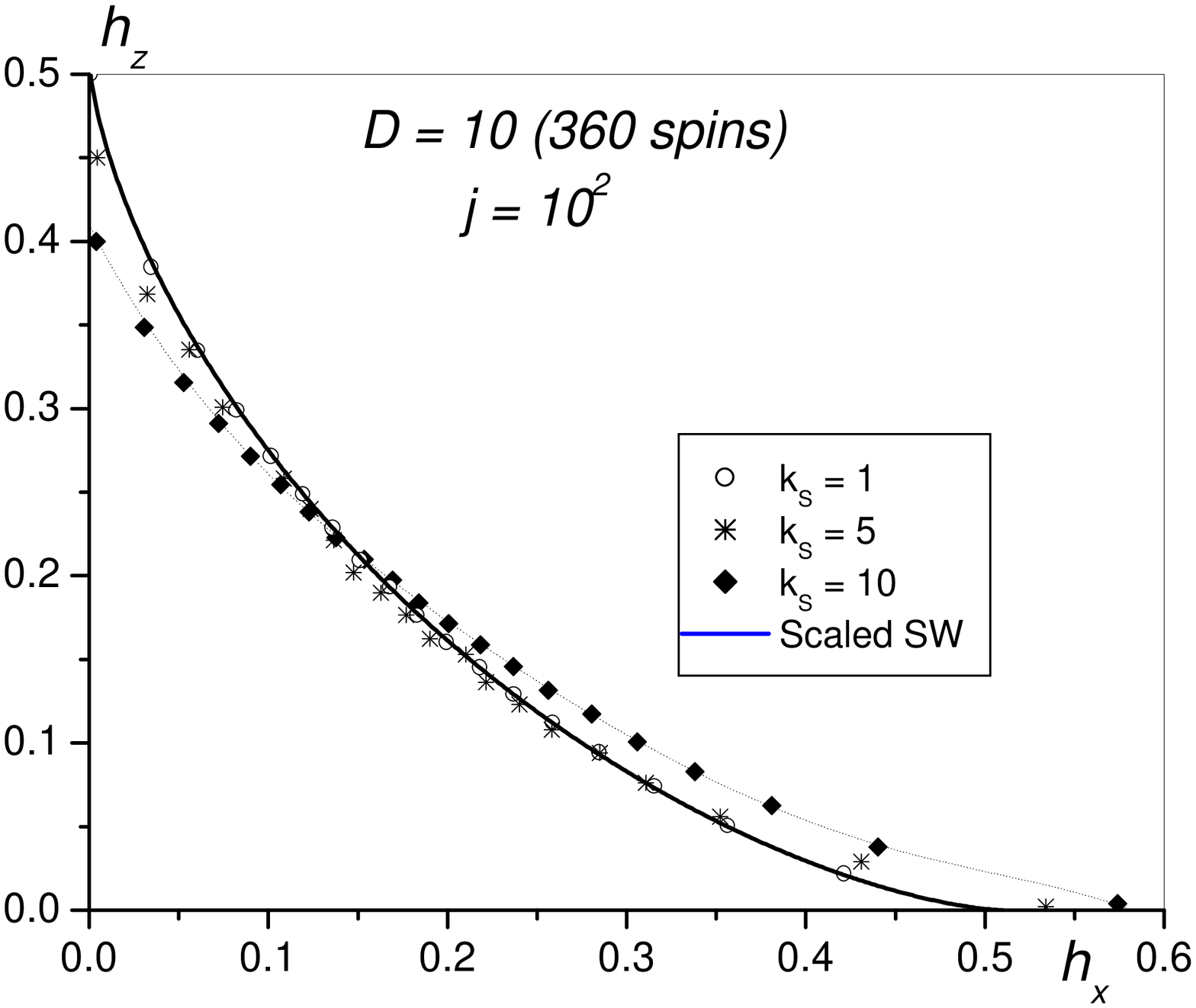}
\caption{\label{astroid_MSP} 
Astroid of a spherical particle of ${\cal N}=360$ spins. 
Left: for $k_{s} = K_s/K_c =1$ and various values of the surface-to-volume ratio $N_{st}\equiv N_s/{\cal N}$.
Right: various values of surface anisotropy constant $k_{s} = K_s/K_c$. The full dark line is the SW astroid scaled with $N_c/{\cal N}$, where $N_c$ is the number of spins in the core; the dotted line is only a guide to the eye.}
\end{figure*}

\begin{figure}[floatfix]
\begin{center}
\includegraphics[width=8cm]{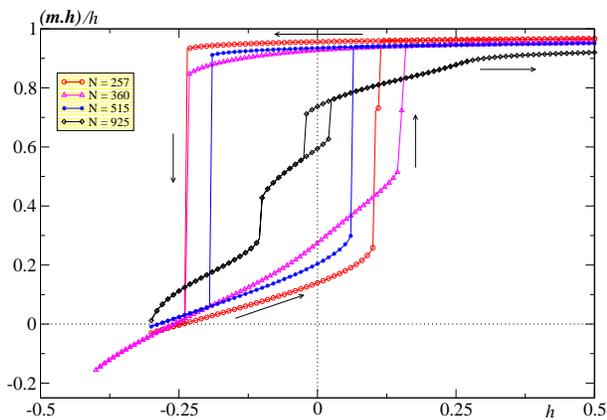}
\end{center}
\caption{Minor hysteresis loops for the particle diameters $D=9,10, 11,13$, which corresponds to the sizes indicated in the legend. $K_c/J = 0.01, K_s/J = 0.3$. The return field is $h=-0.4$ ($h=H/H_k,\,H_k=2K_c$).}
\label{loops_minor_size}
\end{figure}

Fig.~\ref{astroid_MSP} shows the $2D$ limit-of-metastability curve (or SW astroid) for the particle as in Fig.~\ref{hyst_ks_a45}. On the left panel we see that for small surface anisotropy this curve perfectly scales with the SW model (in full line) with the scaling factor $N_c/{\cal N}$, where $N_c$ is the number of core spins, i.e. those spins with full coordination number. On the other hand, the right panel shows that as the surface anisotropy increases this curve depresses in the longitudinal direction, thus deviating from the SW curve.
%
%
It is clear that as $K_s$ increases the spin non-collinearities are enhanced and the SW model is no longer applicable.
In Ref.~\onlinecite{kacgar01physa300} we showed that the spin disorder caused by boundary defects is of long range and propagates to the center of the particle, which is at variance with the so-called core-shell model, as far as the magnetic properties are concerned.
We have also studied minor hysteresis loops as shown in Fig.~\ref{loops_minor_size}. Their appearance is an indication of the existence of several local minima induced by surface anisotropy. Indeed, we found that the bigger is the particle, and thereby the smaller is the surface contribution, and the narrower is the minor loop, which implies that the non-collinearities caused by surface anisotropy are indeed responsible for the minor loop.
%
%

\begin{figure}[floatfix]
\includegraphics[width=8cm]{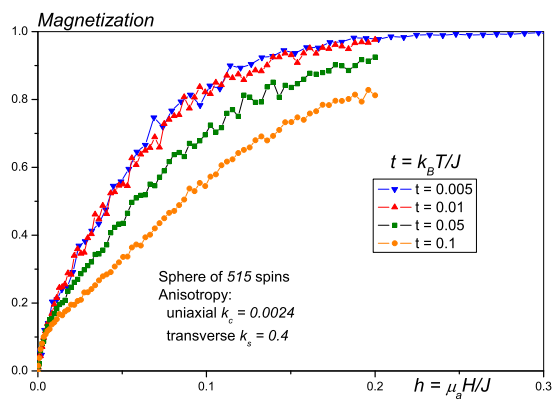}
\caption{Magnetization as a function of field with variable temperature ($t\equiv k_BT/J$) for a particle of 515 spins with core uniaxial anisotropy ($K_c/J=0.0024$) and transverse surface anisotropy ($K_s/J=0.4$).}
\label{SurfaceEffects_MSP}
\end{figure}

At finite temperature, one can study equilibrium thermodynamic properties by Monte Carlo simulations \cite{kacetal00epjb} or by the adapted spherical model \cite{kacgar01physa300} or spin-wave theory \cite{kacgar01epjb}.
For a box-shaped particle with pbc or fbc and without surface anisotropy, we computed the magnetization as a function of field and temperature both analytically and numerically. We used respectively the modified spin-wave theory and Monte Carlo simulation using an augmented Metropolis algorithm that takes accounts of the superparamagnetic global rotation of the whole bunch of spins in the particle.
For instance, Fig.~\ref{SurfaceEffects_MSP} shows the magnetization of a many-spin particle as a function of magnetic field at different temperatures.
%
%
These results obtained by Monte Carlo simulations simply show that because of surface effects the magnetization does not saturate even at relatively high fields, i.e. $h=0.2$ which corresponds to $H\simeq 16$ Tesla for cobalt particles even at very low temperature, i.e. $t=0.01$ ($T \sim 1$ K).
In Ref.~\cite{kacetal00epjb}, it was shown that the long-range order of the spin disorder at the particle's boundary, mentioned above, is accentuated by thermal fluctuations. 

The study of dynamic switching of the magnetization triggered by thermal fluctuations is more involved and one has to resort to stochastic approaches based on the Landau-Lifshitz-Langevin equation, as a dynamic equation for the magnetization, or the Fokker-Planck equation, as a dynamic equation for the probability distribution of the magnetization orientations.
It is also possible to use the so-called kinetic Monte Carlo technique.
On the other hand, the approach based on the solution of the Fokker-Planck equation by the matrix-continued fraction method \cite{cofkalwal05worldsc} is limited to a small number of degrees of freedom, or equivalently, a small number of spins inside the particle. 
The main reason is that with a many-spin particle one is faced with complex many-body aspects with the inherent difficulties related with analysing the energyscape (location of the minima, maxima, and saddle points of the energy potential). This analysis is, however, unavoidable since it is a crucial step in the calculation of the relaxation time and thereby in the study of the magnetization stability against thermally-activated reversal.
On the other hand, the Langevin approach is quite versatile as it can be applied to a general system, but then one has to average over many stochastic trajectories generated with a random Gaussian field. This renders this approach rather time consuming.
For an OSP, the limit-of-metastability curve has been obtained at different temperatures in Ref.~\onlinecite{vouilleetal04jmmm}.
Such thermal effects had also been studied in Ref.~\onlinecite{kacgar01physa291} by constructing a free energy that interpolates between the low-temperature Landau-Lifshitz energy and the high-temperature Landau energy.
It was found that there is a crossover between the low-temperature regime where the magnetization switching follows the SW model, i.e., through coherent rotation, and the high-temperature regime where the magnetization switches by varying its magnitude.
In both works, it is shown that the thermal fluctuations induces a shrinking of the limit-of-metastability curve, which was observed by $\mu$-SQUID measurements \cite{wernsdorfer01acp}.
In the case of a many-spin particle Landau-Lifshitz-Langevin calculations may be performed in order to investigate the effects of both temperature and surface anisotropy on the hysteresis loop and limit-of-metastability curve.

In Ref.~\onlinecite{kacetal00epjb} we investigated, in a somewhat phenomenological way, the effect of varying the ratio of the exchange coupling at the surface to that in the core in maghemite particles. The idea was to model the influence of the matrix in which the particles are embedded. The exchange coupling at the surface is meant to be an effective coupling resulting from the various chemical and physical interactions between the matrix and the particles.
Using the Monte Carlo technique we computed the thermal variation of the surface magnetisation as a function of the (reduced) temperature for a ratio of the exchange coupling at the surface to that in the core $J_{sc}=0.5,1,2$ and for a surface contribution of $N_{st}=40\%$.
We found that the surface critical region is shifted to higher temperatures upon increasing $J_{s}$, and only when $J_{s}=2J_{c}$ do both the core and surface magnetic phase transitions occur in the same
temperature range. The results also showed that the weaker is the exchange interactions on the surface the lower is the magnetisation of the latter.
This result remains the same upon decreasing the outer shell thickness. In this case the number of spins having smaller coordination numbers than in the core, and hence weaker effective exchange energy, i.e., those spins on the outer shell of the particle, is very small as compared to the rest of spins in the particle.
Moreover, we may think that the spins on the outer shell follow the (strong) effective field created by the other (inner) spins constituting a relatively ferrimagnetically ordered core.

\subsection{MSP versus effective OSP}
\begin{figure*}[floatfix]
\includegraphics[width=4cm, angle=-90]{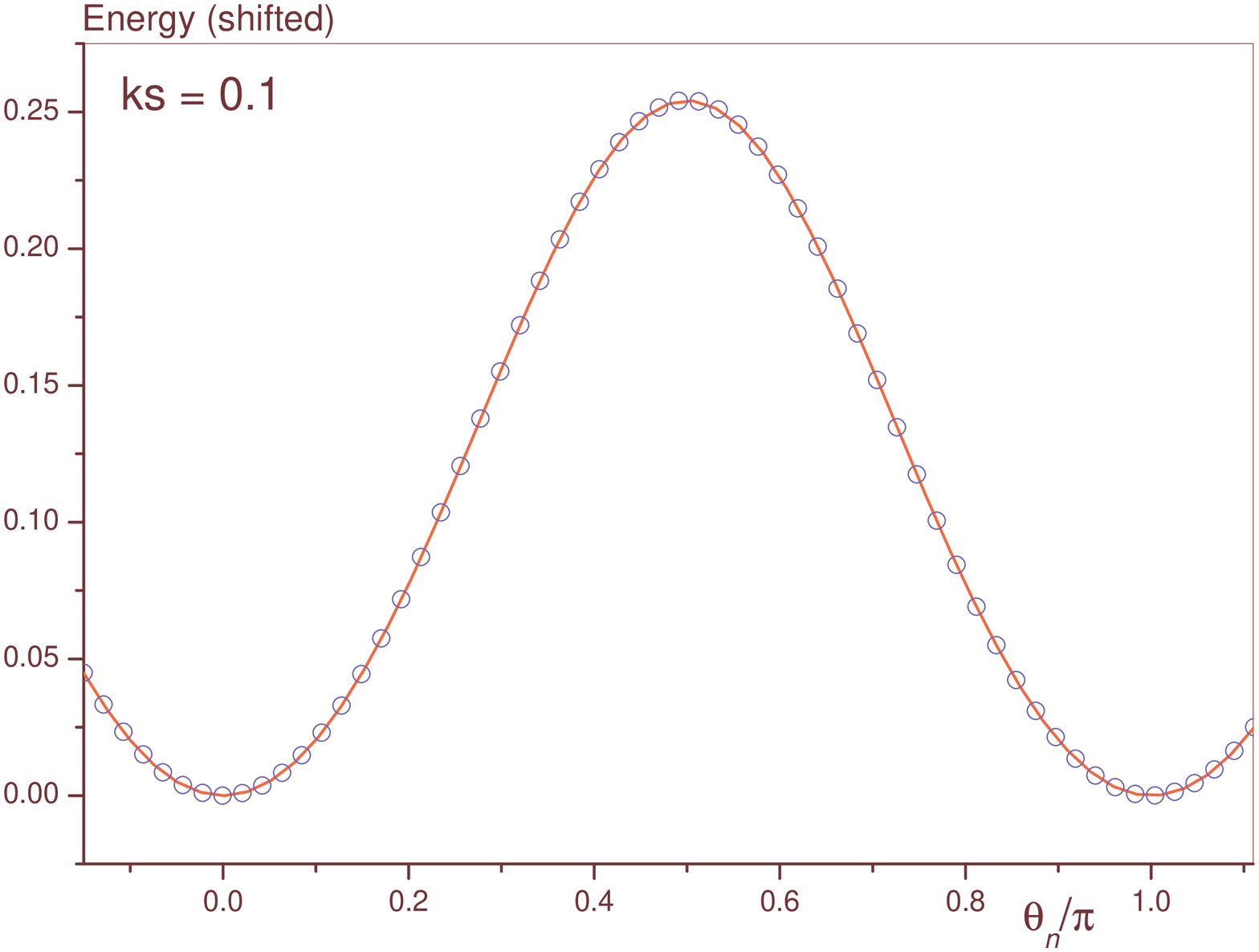}
\includegraphics[width=4cm, angle=-90]{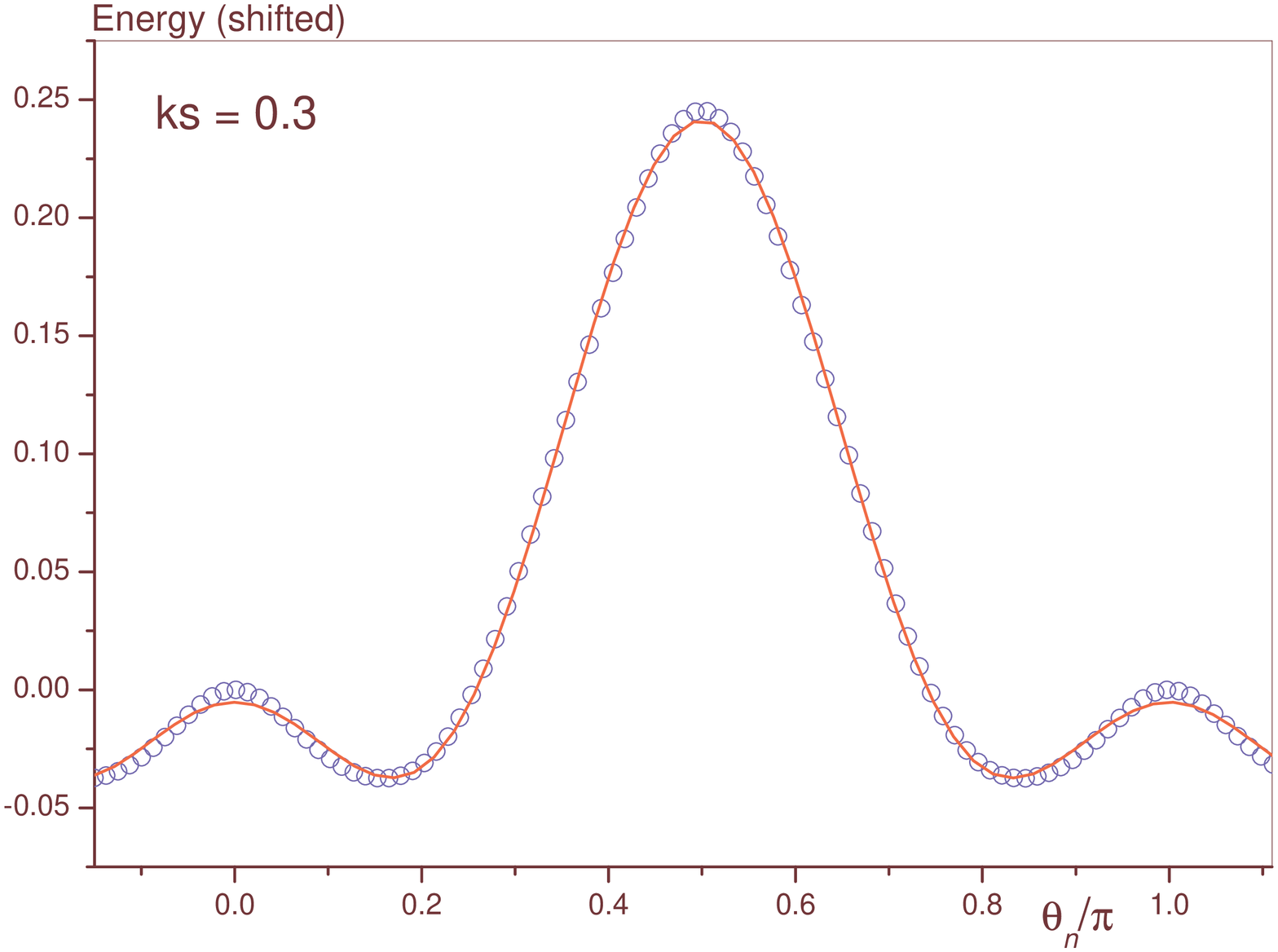}
\includegraphics[width=4cm, angle=-90]{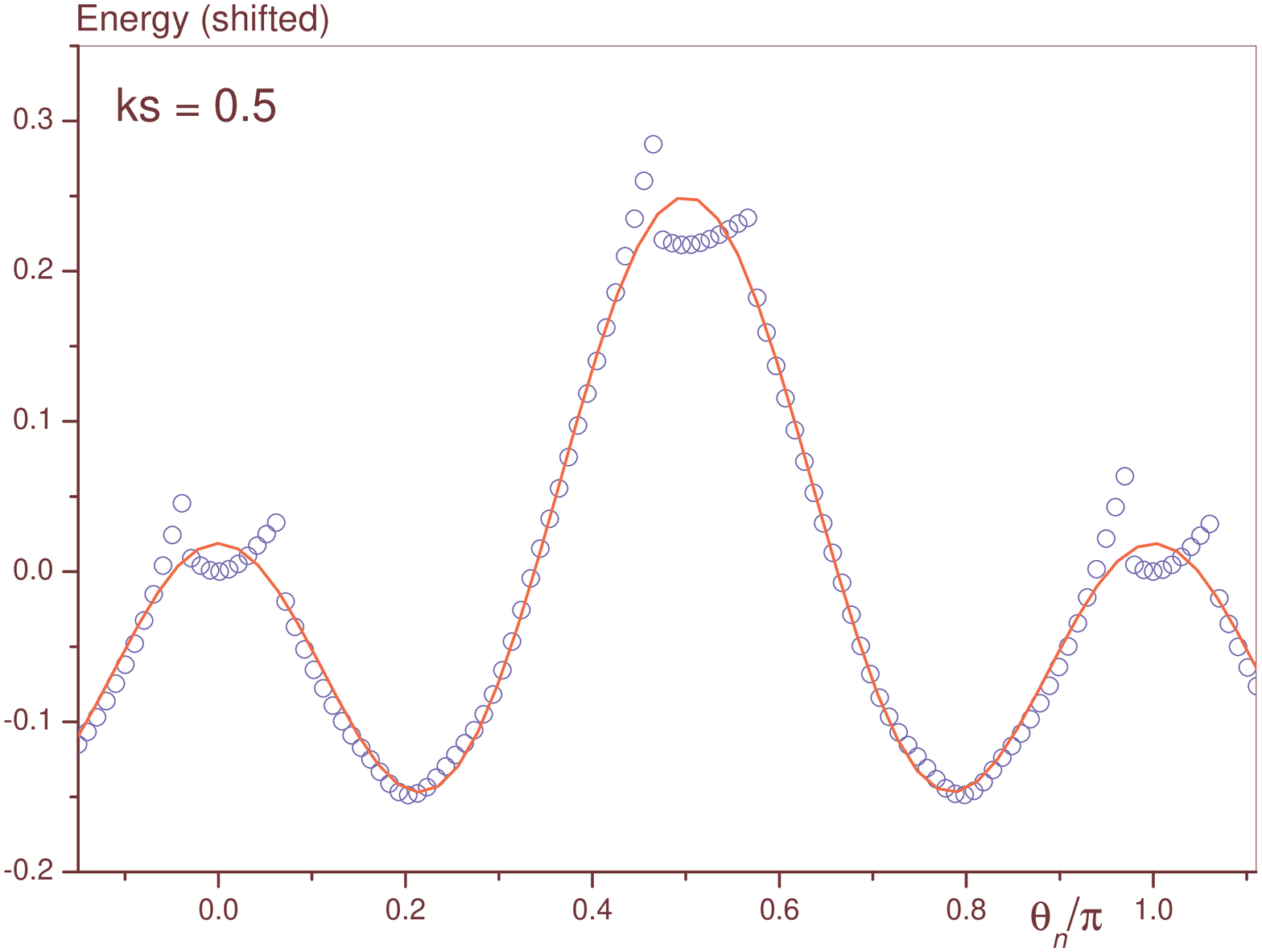}
\caption{(Color online) $2D$ energyscape for the spherical particle of TSA with varying constant $k_s = K_s/J$. The symbols are results of the numerical calculations for the MSP, and the full lines are fits using Eq. (\ref{UniaxialCubicEnergy}).
$\varphi_n=0$ and $h=0$.}
\label{enscape2D_h0_varyks}
\end{figure*}

With the desire to avoid the difficulties mentioned above, one may address the question as to whether there exist some cases in which the full-fledged theory that has been developed for the OSP approach [see \cite{cofkalwal05worldsc} and references therein] can still be used to describe the dynamics of an MSP. However, avoiding somehow the spin non-collinearities induced by surface/interface anisotropy
means that some price has to be paid.
In Ref.~\cite{garkac03prl} it was shown that when the surface anisotropy is much smaller than the exchange interaction, and in the absence of core anisotropy, the surface anisotropy contribution to the particle's energy is of $4^\mathrm{th}$-order in the net magnetization components and $2^\mathrm{sd}$-order in the surface anisotropy constant.
This means that the energy of an MSP, cut from a lattice of cubic symmetry, with relatively weak surface anisotropy can be modeled by that of an OSP whose effective energy contains an additional cubic-anisotropy potential.
An analytical expression was given for the effective constant $K_\mathrm{eff}$ of the surface-induced cubic-anisotropy term, when the core anisotropy is absent, that is
\begin{equation}\label{Keff}
K_\mathrm{eff} = \kappa \frac{\mathcal{N}K_s^2}{z J},
\end{equation}
where $\mathcal{N}, K_s, z, J$ are respectively the number of atoms, the surface anisotropy constant (transverse or N\'eel), the coordination number, and the exchange coupling of the many-spin particle. $\kappa$ is a surface integral that depends on the underlying lattice, the shape, and the size of the particle and also on the surface-anisotropy model. For a spherical particle (of $\sim 1500$ spins) cut from a simple cubic lattice with N\'eel's surface anisotropy, $\kappa\simeq 0.53465$.
In the presence of core anisotropy the effective energy contains terms with coefficients that are products of the core and surface anisotropy constants. However, for small surface anisotropy the effective energy again provides a good approximation of the many-spin particle, as has been shown in Ref.~\cite{kacbon06prb}.
In this case the energy of a many-spin particle with uniaxial anisotropy in the core and TSA or NSA on the surface is modeled by that of a one-spin particle with the net magnetization $\mathbf{m}$ in a potential containing a uniaxial and cubic anisotropy terms, i.e., up to a constant, we have
\begin{equation}\label{UniaxialCubicEnergy}
\mathcal{E}_{\mathrm{eff}} = -K_{\mathrm{uni}}\,m_{z}^{2}+K_{\mathrm{cub}}(m_{x}^{4}+m_{y}^{4}+m_{z}^{4}).
\end{equation}
The results are shown in Fig.~\ref{enscape2D_h0_varyks}.

Before we discuss the results let us briefly explain the method we used to obtain them.
Because we are dealing with an MSP, the energyscape cannot be represented in terms of the coordinates of all spins. Instead, we may represent it in terms of the coordinates of the particle's net magnetization. For this purpose, we fix the global or net magnetization, $\mathbf{m}$, of the particle in a desired direction ${\bm m}_{0}$ ($|{\bm m}_{0}|=1$) by using the energy function with a Lagrange multiplier ${\bm\lambda }$ \cite{garkac03prl}:
\begin{equation}  \label{FFuncDef}
\mathcal{F}=\mathcal{H}-\mathcal{N}{\bm \lambda \cdot }\left( {\bm m}-{\bm m}%
_{0}\right) ,\qquad {\bm m\equiv }\frac{\sum_{i}\mathbf{s}_{i}}{\left|
\sum_{i}\mathbf{s}_{i}\right| }.
\end{equation}
To minimize $\mathcal{F},$ we solve the evolution equations
\begin{eqnarray}\label{LLEqs} 
\mathbf{\dot{s}}_{i} &=&-\left[ \mathbf{s}_{i}\times \left[ \mathbf{s}_{i}\times \mathbf{F}_{i}\right] \right] ,\qquad \mathbf{F}_{i}\equiv
-\partial \mathcal{F}/\partial \mathbf{s}_{i}  \nonumber  \\
{\dot{\bm \lambda }} &=&\mathbf{\partial }\mathcal{F}/\partial {\bm \lambda} = -\mathcal{N}\left( {\bm m}-{\bm m}_{0}\right) ,
\end{eqnarray}
starting from $\mathbf{s}_{i}={\bm m}_{0}=\mathbf{m}$ for all $i=1,\ldots,\mathcal{N}$ and ${\bm\lambda =0}$, until a stationary state is reached. In this state ${\bm m}={\bm m}_{0}$ and $\left[ \mathbf{s}_{i}\times \mathbf{F}_{i}\right] =0,$ i.e., the torque due to the term $\mathcal{N}{\bm\lambda \cdot }\left( {\bm m}-{\bm m}_{0}\right)$ in $\mathcal{F}$ compensates for the torque acting to rotate the global magnetization towards the minimum-energy directions [see discussion in Ref.~\cite{garkac03prl}]. The orientation of the net magnetization is then given either in Cartesian coordinates $(m_{x},m_{y},m_{z})$ or in spherical coordinates $(\theta_{n},\varphi_{n})$.
%
%

\begin{figure}[floatfix]
\includegraphics[width=8cm]{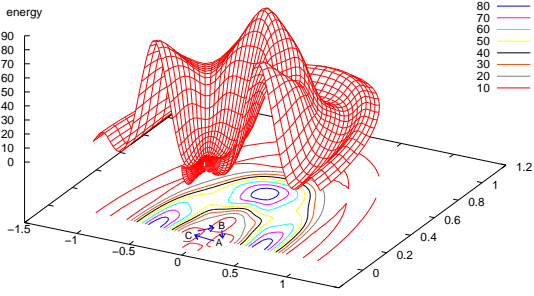}
\caption{Plot of $(\frac{\theta_n}{\pi}\cos\varphi_n ,\frac{\theta_n}{\pi}\sin \varphi_n ,E)$ for the same parameters as before.}
\label{energy_contour_plot}
\end{figure}

\begin{figure}[floatfix]
\includegraphics[width=8cm]{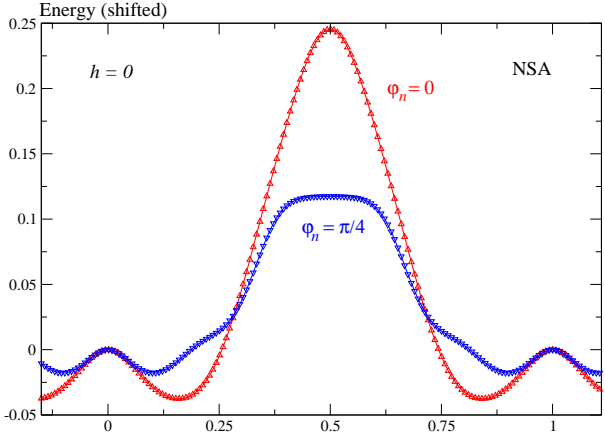}
\caption{(Color online) Energy as a function of the polar angle $\theta_{n}$ for a spherical particle with uniaxial anisotropy in the core and N\'eel anisotropy on the surface. $\varphi_n=0,\pi/4$ and $h=0$.}
\label{NSA360kc001ks03_p0p45}
\end{figure}

The first panel of Fig.~\ref{enscape2D_h0_varyks} ($k_s = K_s/J = 0.1$) shows that for very small $k_s$ the energyscape of an MSP is well recovered by the effective energy in Eq.~(\ref{UniaxialCubicEnergy}).
As $k_s$ increases [see middle panel, $k_s=0.3$], some deviations start to be seen, and for relatively large values of $k_s$ a fit with Eq.~(\ref{UniaxialCubicEnergy}) is no longer possible. In fact, in this regime strong deviations from collinearity develop, especially near maxima and saddle points. In fact, in this case the Lagrange-parameter method of Eqs.~(\ref{FFuncDef}, \ref{LLEqs}) fails because the magnetic state of an MSP can no longer be represented by a net magnetization.
These results imply that the effect of spin non-collinearities on the energy is to split the minimum at $\theta_n=0$, defined by the core uniaxial anisotropy, into four minima at $\theta_n\sim 28^{\circ }$ and $\varphi_n=0,\pm\pi/2, \pi$, reminiscent of cubic anisotropy [see Fig.~\ref{energy_contour_plot}].
These minima are connected by saddle points at $\varphi_n=\pm \pi /4$ and $\pm 3\pi /4$ and the point at $\theta_n=0$ becomes a small local maximum. The four minima exist over a finite range of the applied field, although their positions change continuously as a function of the field \cite{kacbon06prb}.
%
%
Fig.~\ref{NSA360kc001ks03_p0p45} is a plot of the $2D$ energyscape for a spherical particle with uniaxial anisotropy in the core, as before, but now with NSA on the surface.
%
%
It is clear that the cubic-anisotropy features are also seen in the case of NSA, namely that i) the energy minima are not along the directions ($\theta_n = 0,\pi$) of the core easy axis; these directions having become local maxima, as discussed in the case of TSA, and ii) there is a clear dependence on the azimuthal angle $\varphi_n$.
In addition, we would like to mention that more extensive calculations \cite{fesenkoetal06preprint} have shown that similar features are also observed for other crystal structures (fcc, bcc, etc.).

These results agree and complement those of Ref.~\onlinecite{kacdim02prb} and \onlinecite{kacmah04jmmm}, where it was shown that for both the TSA and NSA models there exists a (different) critical value of the surface anisotropy constant that separates i) the OSP SW regime of coherent switching and ii) the MSP regime where the strong spin non-collinearities invalidate the coherent mechanism, and the particle can no longer be modeled by an effective OSP.
Obviously, for very small surface anisotropy the cubic contribution becomes negligible [see the first panel in Fig.~\ref{enscape2D_h0_varyks}, $k_s=0.1$] and thus the SW OSP model provides a good approximation to the many-spin particle.
Accordingly, some experimental macroscopic estimations of the surface anisotropy constant yield, e.g., for cobalt $K_s/J\simeq 0.1$ \cite{skocoe99iop}, for iron $K_s/J\simeq 0.06$ \cite{urquhartetal88jap}, and for maghemite particles $K_s/J\simeq 0.04$ \cite{perrai05springer}.
However, one should not forget that this effective constant depends on the particle's size, among other parameters such as the material composition, and for, e.g. a diameter of $2$ nm we may expect higher anisotropies.

Using the effective energy (\ref{UniaxialCubicEnergy}) we have also studied \cite{kacsch06prb} the ferromagnetic resonance in systems with competing uniaxial and cubic anisotropies and built a model that applies to i) magnetic materials with both uniaxial and cubic anisotropies, and ii) magnetic nanoparticles with core and effective surface anisotropies.
We numerically computed the resonance frequency as a function of the field and the resonance field as a function of the direction of the applied field for an arbitrary ratio of cubic-to-uniaxial anisotropy. 
We also provided some analytical approximate expressions in the case of weak cubic anisotropy and proposed a method that uses these expressions for estimating the uniaxial and cubic anisotropy constants, and for determining the relative orientation of the cubic anisotropy axes with respect to the crystal principle axes.
This method is applicable to the analysis of experimental data of resonance type measurements.

It would very interesting to investigate the crossover between the OSP and MSP regimes, in terms of the particle's size and environment and material's properties. In addition, by comparing the experimental results with various models (e.g., TSA and NSA) for surface anisotropy, one should be able to determine the most appropriate model.

\section{\label{Collective}Collective effects}
While measurements of isolated single particles by the technique of $\mu$-SQUID \cite{wernsdorfer01acp} have been so successful and have provided us with invaluable information on their intrinsic static and dynamic properties, they still suffer from some limitations.
Indeed, apart from the limited range of temperature and applied field, the magnetization value is still inaccessible and only the field at which it changes sign is detectable. This prevents us from obtaining the magnetization as a function of the field, temperature and time, and in particular no hysteresis loop is available.
On the contrary, measurements of particle assemblies are possible in larger ranges of applied field and temperature. However, the analysis of the corresponding results is made rather difficult and subtle because of the volume and easy-axis orientation distributions, and more importantly, the DDI, whose intensity depends on the sample concentration.
At present, most of the studies on magnetic nanoparticles are conducted on assemblies. This is mainly due to the above mentioned reasons and also because today one may understand better how assemblies of nanoparticles can be used in technological applications than one does for an isolated particle.
These applications require ever denser assemblies and thus smaller particles. However, this brings a new dilemma because small particles become superparamagnetic at even low temperature. Moreover, high density entails strong DDI among the particles, and in technological applications such as magnetic recording, this is an issue of special importance because DDI have been widely recognized as being responsible for the deterioration of the signal-to-noise ratio [see e.g., Refs.~\onlinecite{sharrock90ieee, johnson91jap} and references therein]. As such, an optimum material [with appropriate anisotropy and other physical parameters] has still to be devised.
On the other hand, the study of nanoparticle assemblies brings new headaches to theorists, at least, since they are faced with tremendous difficulties related with DDI between particles and their interplay with surface effects intrinsic to the particles.
Nevertheless, the problem of DDI in nanoparticle assemblies has triggered much interest due to many new phenomena that emerge from the collective behavior of the particles, notably the so-called {\it spin-glass state} at low temperature in concentrated assemblies [see Refs.~\cite{doretal99jmmm, troetal00jmmm, jonetal01prb}, and many references therein], and also because these interactions have always constituted a challenging issue in many areas of physics.

Today, there arises the important issue about assemblies of nanoparticles that concerns the understanding of the interplay between the {\it intrinsic} properties, such as those pertaining to surface effects, and {\it extrinsic} or collective
effects stemming from the long-range DDI.
Many research groups have experimentally studied this interplay in cobalt and maghemite particle assemblies.
Measurements of the magnetization at high fields performed on the $\gamma$-Fe$_{2}$O$_{3}$ nanoparticles \cite{ezzir98phd, troncetal00jmmm}, [see also \cite{cheetal95prb} for cobalt particles] have shown that the magnetization is strongly influenced by the particle's size.
For instance, Fig.~1 of Ref.~\cite{kacetal00epjb} shows that i) there is a sudden increase of the magnetization as a function of the applied field when the temperature reaches $70$ K, and the magnetization does not saturate at the
highest available field, i.e., 5.5 T.
ii) there is an important increase of the magnetization at low temperature, iii) the thermal behavior of the magnetization at 5.5 T is such that the smaller is the mean diameter of the particle the faster is the increase of the magnetization at very low temperature.

On the theory side, the situation involving both surface effects and DDI has never been considered so far mainly because of its tremendous complexities, and also because one has first to understand these two effects separately.
Needless to say that, already at the static level, no exact analytical treatment of any kind is ever possible even in the OSP approximation, i.e., ignoring the internal structure of the particles and thereby surface effects.
Only numerical approaches such as the Monte Carlo technique can be of some rescue here. This technique has been used in Ref.~\cite{kectro98prb} to study hysteretic properties of monodisperse assemblies of nanoparticles with the more realistic Heisenberg spin model, in the OSP approximation.
In the case of weak interactions, as is applicable in dilute assemblies, some approximate analytical expressions can be obtained. Indeed, in Ref.~\onlinecite{jongar01prb}, the Landau-Lifshitz thermodynamic perturbation theory~\cite{lanlif80} is used to tackle the case of weakly dipolar-interacting monodisperse assemblies of magnetic moments, i.e., in the OSP approximation, with uniformly or randomly distributed anisotropy axes. The authors studied the influence of DDI on the susceptibility and specific heat of the assembly.
In Ref.~\onlinecite{kacaze05epjb}, the same approach was used with the objective to study the effect of anisotropy and (weak) dipolar interactions on the field and temperature behavior of the magnetization of a monodisperse and polydisperse assemblies of magnetic moments.
We studied an assembly of magnetic moments whose magnitudes are distributed according to a lognormal function. The anisotropy is taken as uniaxial and either textured along some reference axis or randomly distributed. The low-field regime magnetization obtained in Ref.~\onlinecite{raiste02prb}, is generalized so as to take account of polydispersity and DDI. In high fields, the magnetization as a function of temperature and field was computed using the steepest-descent approximation.
In the general range of temperature, field, and anisotropy, the magnetization of a non-interacting assembly was computed exactly by numerical integration of the single-moment (free) partition function. For interacting assemblies, we used the Monte Carlo technique. It was possible to obtain practical (semi) analytical formulae for the field and temperature dependence of the assembly magnetization that take into account moment and easy axes distributions, and weak DDI. It was then possible to investigate the effect of anisotropy and DDI and to discuss the validity of the Langevin law, for both textured and random-anisotropy, which is very often used in the literature to interpret the magnetization measurements on nanoparticle assemblies.

In the following sections, we review the main results of this work and also discuss its extention to include intrinsic effects.
\subsection{Assembly of OSP particles: effects of anisotropy and DDI}
\subsubsection{\label{basics}Notation and basic equations}
We consider an assembly of magnetic moments ${\bf m}_i = m_i{\bf s}_i,\, i=1,\ldots,{\cal N}$ of magnitude $m_i$ and direction ${\bf s}_i$, with $\vert{\bf s}_i\vert=1$. The magnitude of the magnetic moment ${\bf m}_i$ is defined in terms of the Bohr magneton $\mu_B$, i.e., $m_i=n_i\mu_B$, and the numbers $n_i$ are either all equal for monodisperse assemblies or lognormal distributed for  polydisperse assemblies.
Each magnetic moment is assigned a uniaxial easy axis ${\bf e}_i$, and for an assembly these axes are either all directed along some reference axis leading to a textured assembly, or randomly distributed.

The energy of a magnetic moment ${\bf m}_i$ with uniaxial anisotropy axis ${\bf e}_{i}$, interacting with all the other moments via DDI, in the magnetic field ${\bf H}=H{\bf e}_h$, reads [after multiplying by $-\beta=-1/k_BT$],
\begin{widetext}
\begin{eqnarray}\label{Ei_dimless}
{\cal E}_i & = & x_i{\bf s}_i\cdot{\bf e}_{h} + \sigma_i({\bf s}_{i}\cdot{\bf e}_{i})^{2} + \xi_d\sum _{j<i}n_i n_j\frac{3({\bf s}_{i}\cdot{\bf e}_{ij})({\bf s}_{j}\cdot{\bf e}_{ij})-{\bf s}_{i}\cdot{\bf s}_{j}}{r_{ij}^{3}} \\ \nonumber
&\equiv& {\cal E}_i^{(0)} + \xi_d\sum _{j<i}n_i n_j\frac{3({\bf s}_{i}\cdot{\bf e}_{ij})({\bf s}_{j}\cdot{\bf e}_{ij})-{\bf s}_{i}\cdot{\bf s}_{j}}{r_{ij}^{3}}.
\end{eqnarray}
\end{widetext}
where
\begin{equation}\label{Ed_rij}
{\bf r}_{ij}={\bf r}_{i}-{\bf r}_{j},\, \, \, {\bf e}_{ij}={\bf r}_{ij}/r_{ij}
\end{equation}
is the vector joining the sites $i$ and $j$ and whose magnitude is measured in units of $a$, a characteristic length on the lattice to be specified later.
${\cal E}_i^{(0)}$ is the free particle energy, and
\begin{equation}\label{dimless_params}
x = \frac{\mu_BH}{k_BT}, \quad \sigma = \frac{\mu_BK}{M_s k_BT}, \quad \xi_d = \frac{\mu_0 \mu_B^2}{4 \pi a^{3}k_{B}T},
\end{equation}
are the dimensionless energy parameters. It is also convenient to introduce the parameters $x_i=xn_i, \sigma_i=\sigma n_i$. Note that $\sigma_i=KV_i/(k_BT)$ is the commonly used notation for the reduced anisotropy-barrier height of the particle $i$.

The magnetization per particle of the assembly, or its component in the field direction, which is taken here along the $z$ axis, is given by
\begin{equation}\label{assembly_mag}
\left\langle m^{z}_{as}\right\rangle (\sigma ,x, \xi_d)=\frac{1}{\cal N}\int
\frac{d^{2}e_{i}}{2\pi }\sum_{i=1}^{\cal N}w(n_i)\, \left\langle
m^{z}_{i}\right\rangle (m_{i},{\bf e}_{i},\sigma ,x, \xi_d),
\end{equation}
where $w(n_i)$ is the lognormal distribution of the numbers $n_i$ with parameters $\mu, \delta$,
\begin{equation}\label{lognormal}
w(n)=\frac{1}{n\delta\sqrt{2\pi}}\exp \left[-\frac{1}{2}\left(\frac{\ln n-\mu }
{\delta }\right) ^{2}\right].
\end{equation}
In Eq.~(\ref{assembly_mag}), $\left\langle m^{z}_{i}\right\rangle$ is the statistical thermodynamic average  over the particle's moment direction ${\bf s}_i$,
\begin{equation}\label{direction_average}
\left\langle m^{z}_{i}\right\rangle =\frac{1}{Z}\int {\cal D}\Omega\,
e^{\cal E}\, m^{z}_{i}=m_{i}\left\langle s_{i}^{z}\right\rangle, \qquad Z=\int D\Omega \, e^{\cal E},
\end{equation}
where  ${\cal D}\Omega =\prod _{i}d\Omega _{i}=\prod _{i}d^{2}s_{i}/2\pi$, and ${\cal E}=\sum_i{\cal E}_i$.

\subsubsection{Effect of anisotropy and DDI on $M(T,H)$}
Applying the thermodynamic perturbation theory \cite{lanlif80} and noting that i) since the field is applied along the $z$ axis the average of the $x,y$ components vanishes, ii) DDI only involve pairs of distinct indices, we obtain the following expression for the magnetization of a weakly interacting assembly (to first order in $\xi_d$) \cite{kacaze05epjb},
\begin{equation}\label{mz_final}
\left\langle S^{z}_{i}\right\rangle \simeq \langle S_{i}^{z}\rangle_0 +
\xi_d\sum_{k=1}^{\mathcal{N}} \langle S_{k}^{z}\rangle_0 A_{ki}
\frac{\partial<S_{i}^{z}>_{0}}{\partial x},
\end{equation}
where ${\bf S}_i = n_i {\bf s}_i$ and
\begin{eqnarray}\label{DDITensors}
 A_{kl} &=& \frac{\left[ 3({\bf e}_{h}\cdot{\bf e}_{kl})^{2}-1\right]}{r_{kl}^{3}}
= {\bf e}_h\cdot{\cal D}_{kl}\cdot{\bf e}_h, \nonumber \\
{\cal D}_{ij} &\equiv& \frac{1}{r_{ij}^3}\left(3{\bf e}_{ij}{\bf e}_{ij}-1\right),
\end{eqnarray}
are the DDI tensors.
As was discussed in Ref.~\onlinecite{jongar01prb} and confirmed in Ref.~\onlinecite{kacaze05epjb}, for non spherical systems, the corrections to the magnetization are largely dominated by the first order contribution to the DDI.

The simple equation (\ref{mz_final}) tells us that the magnetization of a (weakly) interacting particle is given in terms of the magnetization $\langle S_{i}^{z}\rangle_0$ of the free particle and its derivatives with respect to the field. $\langle S_{i}^{z}\rangle_0$ is the statistical thermodynamic average in the absence of DDI and is thus given by Eq.~(\ref{direction_average}) with ${\cal E}_i$ replaced  by ${\cal E}_i^{(0)}$ [see Eq.~(\ref{Ei_dimless})].

Now, the free particle magnetization can be computed either exactly by numerical integration in (\ref{direction_average}) or analytically in some limitting cases, such as low and high field.
Approximate analytical expressions can be obtained in the low-field regime by perturbation theory \cite{garpal00acp, raiste02prb, kacaze05epjb} and in the high-field regime using the steepest-descent approximation \cite{kacaze05epjb}.
More precisely, in low fields we have 
\begin{widetext}
\begin{equation}\label{averaged_lf}
\displaystyle\left\langle s_i^z \right\rangle^\mathrm{lf}_0\simeq
\displaystyle\left\lbrace
\begin{array}{ll}
\displaystyle\frac{1+2S_{i2}}{3}x_i - \frac{7+70S_{i2}^2 + 40S_{i2}-12S_{i4}}{315}x_i^3, &\mathrm{textured} \\ \\
\displaystyle\frac{x_i}{3}-\frac{1+2S_{i2}^2}{45}x_i^3, &\mathrm{random},
\end{array}
\right.
\end{equation}
\end{widetext}
where \cite{raiste02prb}
\begin{equation}\label{S_l_1}
S_{il}(\sigma_i)\simeq
\displaystyle\left\lbrace
\begin{array}{ll}
\frac{(l-1)!!}{(2l+1)!!}(\frac{\sigma_i}{2})^{l/2}+\ldots, &\sigma_i \ll 1, \\ \\
1 - \frac{l(l+1)}{4\sigma_i}+\ldots, &\sigma_i \gg 1.
\end{array}
\right.
\end{equation}
The high-field magnetization reads \cite{kacaze05epjb}
\begin{equation}\label{averaged_sda}
\left\langle s_i^z \right\rangle^\mathrm{hf}_0\simeq
\displaystyle\left\lbrace
\begin{array}{ll}
\displaystyle 1-\frac{1}{x_i}+\frac{\sigma_i}{x_i^2}, &\mathrm{textured} \\ \\
\displaystyle 1-\frac{1}{x_i}-\frac{\sigma_i^2}{15}\frac{1}{x_i^2}, &\mathrm{random}.
\end{array}
\right.
\end{equation}

The validity of the asymptotic low field and high field expressions of the magnetization given in (\ref{averaged_lf}) and (\ref{averaged_sda}), respectively, is checked by comparing them with the exact numerical calculation of the partition function in (\ref{direction_average}).
We find that these asymptotic expressions are good enough even at a relatively high temperature (here small $\sigma$), for which the steepest-descent approximation is expected to work worst because $x$ becomes small.
More general expressions were obtained in Ref.~\onlinecite{garpal00acp} by a different approach.

Next, we generalized the expressions analogous to Eqs.~(\ref{averaged_lf}) and (\ref{averaged_sda}) to a weakly interacting polydisperse assembly in the case of random anisotropy.
Accordingly, inserting the low and high field expansions (\ref{averaged_lf}), (\ref{averaged_sda}) in Eq.~(\ref{mz_final}) leads to analytical expressions for the magnetization of a weakly interacting assembly as a function of field, temperature, anisotropy, and the DDI parameter $\xi_d$.
Therefore, the assembly (reduced) magnetization per particle defined as $\left\langle s_z \right\rangle_\mathrm{ass}=1/{\cal N}\sum_{i=1}^{\cal N}\left\langle s_z \right\rangle_i$, reads (in the case of randomly distributed easy axes)
\begin{widetext}
\begin{equation}\label{lfhf_ddi_mag}
\left\langle s_z \right\rangle_\mathrm{ass}\simeq
\displaystyle\left\lbrace
\begin{array}{ll}
\left[1 + \dfrac{\tilde{\xi}_d}{3}{\cal C}^{(1,2)}\right]\dfrac{\left\langle x\right\rangle}{3}
-\left[A_3 + \dfrac{4}{3}\tilde{\xi}_d A_5 \right]\dfrac{\left\langle x\right\rangle^3}{45} , &\mathrm{low\,field} \\ \\
1-\dfrac{1}{\left\langle x\right\rangle}-\left[\dfrac{\left\langle\sigma\right\rangle^2}{15}-\tilde{\xi}_d{\cal C}^{(0,1)}\right]\dfrac{1}{\left\langle x\right\rangle ^2} + \tilde{\xi}_d\left[\dfrac{2\left\langle\sigma\right\rangle^2}{15}{\cal C}^{(1,1)} - {\cal C}^{(0,0)}\right]\dfrac{1}{\left\langle x\right\rangle ^3}, &\mathrm{high\,field},
\end{array}
\right.
\end{equation}
\end{widetext}
where we have defined [see Eq.~(\ref{dimless_params}) for notation]
$
\tilde{\xi}_d \equiv \xi_d\left\langle n\right\rangle^2, \left\langle x\right\rangle\equiv \left\langle n\right\rangle x, \left\langle \sigma\right\rangle\equiv \left\langle n\right\rangle \sigma$, with $\left\langle n \right\rangle\equiv 1/{\cal N}\sum_{i=1}^{\cal N}n_i,$
and introduced the (scaled) assembly-on-lattice constants
\begin{widetext}
\begin{eqnarray}\label{latticeconstants}
{\cal C}^{(a,b)} = \dfrac{1}{\cal N}\sum_{i,j=1, i\neq j}^{\cal N}\dfrac{n_i^aA_{ij}n_j^b}{\left\langle n\right\rangle^{a+b}}, \quad
A_3 =\dfrac{1}{\cal N}\sum_{i=1}^{\cal N}\dfrac{n_i^3 \alpha_i}{\left\langle n \right\rangle^3}, \quad
A_5 = \dfrac{1}{\cal N}\sum_{i,j=1, i\neq j}^{\cal N}\dfrac{\alpha_i n_i^3 A_{ij}n_j^2}{\left\langle n \right\rangle^5}.
\end{eqnarray}
\end{widetext}
$\alpha_i=1+2S_{i2}^2$ with $S_{i2}$ being defined in (\ref{S_l_1}).

In Eq.~(\ref{lfhf_ddi_mag}), we observe that there are three types of contributions: There are pure anisotropy terms, pure DDI terms, and mixed terms.
It is readily seen that in the absence of anisotropy and DDI, i.e., for $\sigma_i=0,\xi_d=0$, and from Eq.~(\ref{S_l_1}) $\alpha_i=1$, expressions (\ref{lfhf_ddi_mag}) simplify back to Eqs.~(\ref{averaged_lf}), (\ref{averaged_sda}).
The presence of mixed terms simply reminds us that DDI induce an additional anisotropy in the magnetic system and thus re-normalize the pure magneto-crystalline anisotropy.

In the continuum limit the lattice sum ${\cal C}^{(0,0)}=\dfrac{1}{\cal N}\sum_{i,j=1, i\neq j}^{\cal N}A_{ij}$ becomes \cite{jongar01prb} ${\cal C}^{(0,0)} = 4\pi (1/3-\lambda _z)$, for a simple cubic lattice, with $\lambda_z$ being the demagnetizing factor along the $z$ axis. In the monodisperse case we have ${\cal C}^{(a,b)}\propto {\cal C}^{(0,0)}$.
The constants ${\cal C}^{(a,b)}$, and thereby the corresponding DDI terms in Eq.~(\ref{lfhf_ddi_mag}), are shape dependent, which is no surprise knowing that the long range DDI lead to shape dependence of the physical quantities, and in particular the magnetization.
On the other hand, we found that these constants are negative for the oblate system, and positive for the prolate, which implies that the DDI suppress the assembly magnetization in the former case and enhance it in the latter.
Moreover, for cubic systems all the constants ${\cal C}^{(a,b)}$ vanish, which means that the DDI do not contribute to the magnetization in this case (see Fig.~4 of Ref.~\onlinecite{kacaze05epjb}), and thus the deviations from the Langevin behavior are caused only by anisotropy.
Note that all the scaled constants ${\cal C}^{(a,b)}$ are almost independent of the assembly mean diameter. In particular, this is trivial for ${\cal C}^{(0,0)}$.
The constants $A_3,A_5$ contain the parameter $\alpha_i$ and thereby depend on anisotropy.
From Eq.~(\ref{S_l_1}) we infer that in the absence of anisotropy, i.e., $\sigma_i=0$, $\alpha_i=1$, while for strong anisotropy we may approximate $S_{i2}$ to $1$, and hence $\alpha_i$ to $3$, so that $A_3\propto (1/{\cal N})\sum_i n_i^3/\left\langle n \right\rangle^3$ and $A_5\propto{\cal C}^{(2,3)}$ in the two limits of anisotropy. For $D_m=3,7$ nm, $A_3\simeq 2,6$. In the continuum limit $A_3$ tends to $\exp(3\delta^2)$, where $\delta$ is the standard deviation of the distribution (\ref{lognormal}). $A_5$ exhibits the same behavior as ${\cal C}^{(1,2)}$ but with bigger change with $D_m$. It is well known from other areas of physics that the calculation of such high-order moments (or ``cumulants") requires more precision because they present more statistical fluctuations with the lattice size.

\begin{figure*}[ht!]
\includegraphics[width=8cm, angle=-90]{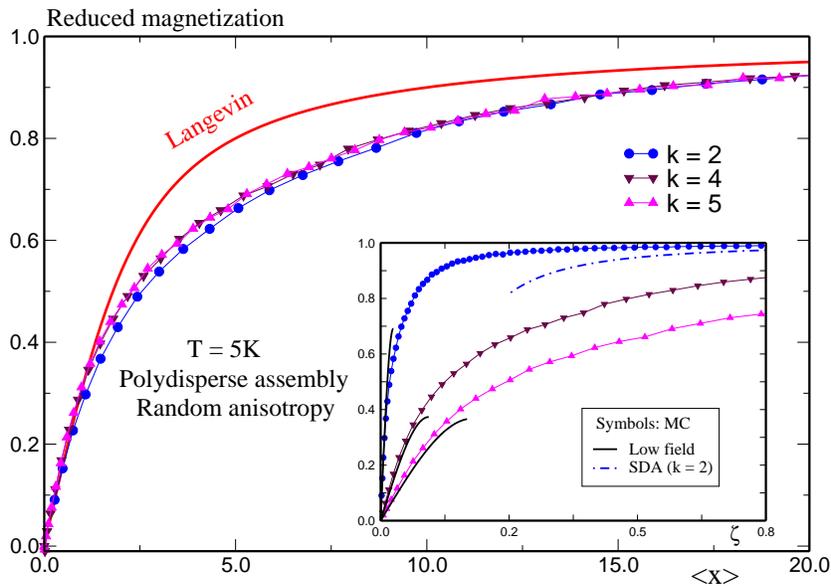}
\caption{\label{osp_ass_ddi_mc_sda}
Reduced magnetization (per particle) of an interacting assembly of ${\cal N} = 10\times 10 \times 5$ lognormal-distributed magnetic moments with mean diameter $Dm = 7 \mathrm{nm}$ and random anisotropy.  Monte-Carlo in symbols and analytical expressions (\ref{lfhf_ddi_mag}) in lines. The parameters $\zeta$ and $k$ are defined in the text.}
\end{figure*}

In Fig.~\ref{osp_ass_ddi_mc_sda} we plot the Langevin function (full line) and the Monte Carlo results (symbols) for the magnetization of an interacting assembly of (${\cal N} = 10\times 10 \times 5$) lognormal-distributed moments, with random anisotropy, and for three values of the inter-particle distance.
The intensity of DDI, or equivalently the value of $\xi_d$, is varied by varying the lattice parameter $a$ entering $\xi_d$ [see Eq.~(\ref{dimless_params})].
More precisely, the parameter $a$ is taken as a real number times the mean diameter $D_m$ of the assembly, i.e., $a=k\times D_m$. Thus, large values of $k$ correspond to an isotropically inflated lattice with large distances between the magnetic moments, and thereby weaker DDI.
These results, obtained for an assembly on a simple cubic lattice, do confirm that DDI suppress the magnetization. Indeed, we recall that it was shown by Luttinger and Tisza \cite{luttis46pr} [see also the more recent work \cite{kectro98prb} using the Monte Carlo technique] that the ground state of a simple cubic lattice of dipoles is antiferromagnetic, while that of a face-centered cubic lattice is ferromagnetic.
%
%
It is also seen that these curves deviate from the Langevin law. However, we emphasize that the deviations induced by DDI are much smaller than those induced by anisotropy, as already discussed earlier [see also Ref.~\onlinecite{jongar01prb} for a related discussion of the effect of the system shape on the magnetic susceptibility of a monodisperse assembly].
In the inset of Fig.~\ref{osp_ass_ddi_mc_sda} the same results are magnified by plotting them in function of $\zeta = (\left\langle x \right\rangle/\tilde{\xi}_d)\times 10^{-3}\propto \mu_BH/(\mu_B^2/a^3)$, i.e., the ratio of Zeeman energy to the DDI energy, which also makes it possible to distinctively plot the analytical expressions (\ref{lfhf_ddi_mag}) for low fields. In the case of high fields only one curve ($k=2$, i.e., relatively strong DDI) is presented since for the other values of DDI parameter $k$, the steepest-descent approximation is only valid for much higher values of $\zeta$.
Note also that in function of the parameter $\zeta$ the tendency with increasing DDI strength is reversed.

\subsubsection{Effect of DDI on the ZFC magnetization of an OSP assembly}
To analyze and understand the dynamics experiments on interacting nanoparticle assemblies, and in particular to understand the system's dynamic response such as the ac susceptibility and zero-field-cooled (ZFC) and field-cooled (FC) magnetization, we need to know how the DDI affect the relaxation time of the assembly.
While a fair understanding of the mechanisms behind the ZFC-FC magnetization and  ac susceptibility has been achieved in the case of non-interacting assemblies, many experimental results on interacting assemblies remain unexplained, notably the observation, mentioned in the introduction to section \ref{Collective}, of a glass-like state at low temperature in strongly interacting assemblies.
Obviously, this is mainly due to the complexities of the DDI and also of the very calculation of the relaxation time itself.
Recently, J\"onsson and Garcia-Palacios \cite{jongar01epl} (see also \cite{garpalgar04prb}) have established an approximate expression for the relaxation rate of a weakly interacting monodisperse assembly with textured or randomly-distributed anisotropy.
Then using the simple Debey relaxation model they investigated the effect of weak DDI on the ac susceptibility and in particular the displacement of the maximum of its real and imaginary parts.
In their explanation of the results they emphasized the important role played by damping in the relaxation processes in the presence of a transverse field, in addition to the effect of the change in the energy barriers, that was commonly believed to play the major role. The role of a transverse field is played here by the transverse component of the dipolar field.

Experimental results obtained for ferrofluids \cite{luoetal91prl} and later for $\gamma$-Fe$_{2}$O$_{3}$ nanoparticles \cite{sapetal97prb} indicated that for dilute samples (weak DDI), the temperature $T_{\max}$ at the maximum of ZFC magnetization first increases with increasing field, attains a maximum and then decreases. More experiments performed on the $\gamma$-Fe$_{2}$O$_{3}$ particles dispersed in polymer \cite{ezzir98phd} confirmed the previous results for dilute samples and showed that, on the contrary, for concentrated samples (strong DDI) $T_{\max }$ was a monotonic decreasing function of the magnetic field.
In Ref.~\onlinecite{kacetal00jpcm} we showed that the bell-like shape of $T_{\max}(H)$, in the dilute case, is insensitive to the intrinsic properties of the particles, of course in the OSP approximation.
Exact numerical calculations \cite{geoetal97acp, coffeyetal98prb, kennedy97phd} of the smallest eigenvalue of the Fokker-Planck matrix invariably led to a monotonic decrease in the blocking temperature, and thereby in the temperature $T_{\max }$, as a function of the magnetic field.
Indeed, in Ref.~\onlinecite{kacetal00jpcm} we showed that the expression of the single-particle relaxation time does not seem to play a crucial role and that even the (relatively) simple N\'eel-Brown expression for the relaxation time leads to a maximum in $T_{\max }(H)$ if one considers an assembly of particles whose magnetization, formulated through the Gittleman-Abeles-Bozowski model \cite{gitetal74prb}, has a superparamagnetic contribution that is a non-linear function (such as Langevin's) of the magnetic field. The magneto-crystalline anisotropy and the volume-distribution width too have strong influence.
The issue of the effect of DDI on $T_{\max }(H)$, namely the disappearance of the maximum when the intensity of interactions increases, was left open in Ref.~\onlinecite{kacetal00jpcm}.
In a recent work \cite{azekac06prep} we revisited this issue after generalizing the work of Ref.~\onlinecite{jongar01epl} to include the magnetic field and magnetic-moment distribution (polydisperse assembly) of nanoparticles in the OSP approximation.
We then investigated the effect of (weak) DDI on the ZFC magnetization, in particular on $T_{\max }(H)$. Below, we briefly outline the main results \cite{azekac06prep}.

The idea is to introduce the effective field ${\bf \xi}$ composed of the external magnetic field and the dipolar field (remember that ${\cal D}_{ii}=0$ and see notation in section \ref{basics}),
\begin{equation}\label{eff_field}
{\bf\zeta}_i = x_i{\bf e}_h + \xi_i^\mathrm{dip}= x_i{\bf e}_h + \xi_d n_i\sum_j {\cal D}_{ij}{\bf S}_{j}.
\end{equation}  

Then the relaxation rate of the assembly is computed as the average of the relaxation rates of each magnetic moment that experiences this effective field using perturbation theory assuming that $\mid{\bf\zeta}_i\mid\ll 1$.
In Ref.~\onlinecite{luisetal04jpcm} an estimation of DDI is given for two samples of Cobalt nanoparticles which indicates that the DDI field is of the order of $300$ Oe, which in reduced units, obtained after dividing by the corresponding anisotropy field of the order of $0.3$ T, is $\xi^\mathrm{ddi}\sim 4\times10^{-3}-10^{-2}$. This is of course very small.
On the other hand, the field at which $T_{\max }(H)$ has a maximum is circa $100$ Oe [see Ref.~\onlinecite{kacetal00jpcm}, Fig. 1, for maghemite particles], corresponding to a reduced field of $3\times 10^{-2}$. This indeed suggests that in typical samples the above condition on the effective field $\zeta $ is satisfied.

The relaxation rate of a weakly DDI-interacting nanomagnet obtained in Ref.~\onlinecite{jongar01epl} is written as
\begin{equation}\label{jongar_rr}
\Gamma_i \simeq  \Gamma_i^{(0)}\left[ 1+\frac{1}{2}\xi _{i,\parallel}^{2}+\frac{1}{4}F(\alpha _{i})\xi _{i,\perp }^{2}\right] ,
\end{equation}
where $\Gamma_i^{(0)}$ is the relaxation rate of the nanomagnet at site $i$ in the absence of the DDI field $\xi_i$ and is given by
\begin{equation}\label{RelaxationRate0}
\Gamma_i^{(0)}=\frac{2}{\tau_s\sqrt{\pi }}\sigma_i^{1/2}\,e^{-\sigma_i},
\end{equation}
with $\tau _{s}=\left(\gamma H_a\right) ^{-1}$ which evaluates to $\gamma\simeq 1.76\times 10^{11}$ (T. s)$^{-1}$ for cobalt particles with the anisotropy field $H_a\sim 0.3$ T; $\gamma\simeq 1.76\times 10^{11}$ (T. s)$^{-1}$.
In the high-energy barrier approximation, $\sigma \gg 1,h=x/2\sigma \ll 1$, which we will be using later on, the function $F$ reads \cite{garetal99pre}
\begin{equation}\label{Fiexpanded}
F(\alpha _{i})\simeq 1-\frac{5}{4\lambda ^{2}}\frac{1}{\sigma _{i}},
\end{equation}
where $\lambda$ is the Landau-Lifshitz damping parameter.
The field $\xi $ in Eq.~(\ref{jongar_rr}) was then replaced by the average over the assembly index and spin orientations of the dipolar field.

In Ref.~\onlinecite{azekac06prep}, we replace the field ${\bf \xi}_i$ by the effective field ${\bf\zeta}_i$ in Eq.~(\ref{eff_field}) and then compute the statistical average over the spin orientation and the (random) anisotropy easy-axis distribution and obtain the double averages
\[
\left\{ 
\begin{array}{l}
\overline{\left\langle \zeta _{i,\parallel }^{2}\right\rangle _{0}}=\frac{1}{3}x_{i}^{2}+\frac{\xi _{d}^{2}}{3}n_{i}^{2}\,R_{ij}, \\
\overline{\left\langle \zeta _{i,\perp }^{2}\right\rangle _{0}}=\frac{2}{3}x_{i}^{2}+\frac{\xi _{d}^{2}}{3}n_{i}^{2}\,\left( 2R_{ij}\right) =2\overline{\left\langle \zeta _{i,\parallel }^{2}\right\rangle _{0}}.
\end{array}
\right. 
\]
where
\[
R_{ij}\equiv \sum\limits_{j\neq i}\frac{2n_{j}^{2}}{r_{ij}^{6}}. 
\]
is a lattice sum.

Therefore, the relaxation rate of a weakly interacting particle containing $n_i$ Bohr magnetons, embedded in a polydisperse assembly, is given by
\begin{equation} \label{DDIRR}
\Gamma_i \simeq  \Gamma_i^{(0)}
\left[ 1 + \frac{1+F(\alpha_i)}{2}
\left(
    \frac{1}{3} x_i^2 + \frac{\xi_d}{2}{3} n_i^2\,R_{ij}
\right)
\right] .
\end{equation}

We then compute the ZFC magnetization of an assembly of magnetic moments $\mathbf{m}_{i}=n_i\mu_B\mathbf{s}_{i},\,i=1,\ldots ,\mathcal{N}$.
For this purpose we use the model introduced by Gittleman et al. \cite{gitetal74prb} which we summarize now.

The ac susceptibility of an assembly of non-interacting particles, with a volume distribution and randomly distributed easy axes, can be written as
\begin{equation}\label{Susc1}
\chi (T,\omega )=\int\limits_0^{\infty}dVf(V)\,\chi_V(T,\omega ),
\end{equation}
where $f(V)$ is the (normalized) lognormal volume distribution with parameters $V_{0}$ and $\delta $
\[
\mathcal{D}V\equiv f(V)dV=\frac{1}{\delta \sqrt{2\pi }}\exp \left[ -\frac{\log ^{2}(\frac{V}{V_0})}{2\delta ^2}\right] \frac{dV}{V}.
\]
$\chi_V$ is calculated by assuming a step function for the magnetic field, i.e. $H=0$ for $t<0$ and $H=H_0=\mathrm{const}$ for $t>0.$ Then, the contribution to the magnetization from particles of volume $V$ is given by
\begin{equation}\label{Susc2}
M_V(t)=VH_0\left[ \chi_0 - (\chi_0-\chi_1)e^{-t/\tau}\right],
\end{equation}
where $\chi_0$ is the susceptibility at thermodynamic equilibrium and $\chi_1$ is the initial susceptibility of particles in the blocked state [see \cite{doretal97acp} and many references therein]. The Fourier transform of (\ref{Susc2}) leads to the complex susceptibility
\begin{equation}\label{Susc3}
\chi =\frac{(\chi_0 + i\omega \tau \chi_1)}{1+i\omega \tau},
\end{equation}
with the real part \cite{gitetal74prb}
\begin{equation}\label{Susc4}
\chi ^{\prime }=\frac{\chi_0+\omega^2\tau^2\chi_1}{1+\omega^2\tau^2},  
\end{equation}
where $\omega$ is the angular frequency ($=2\pi \nu $).
Starting from (\ref{Susc4}) the application of an alternating field yields: a) $\chi^\prime = \chi_0$ if $\omega \tau \ll 1$: At high temperature the magnetic moments orientate themselves on a great number of occasions during the time of a measurement, and thus the susceptibility is the superparamagnetic susceptibility $\chi_0$. b) $\chi^\prime = \chi_1$ if $\omega \tau \gg 1$: At low temperature the energy supplied by the field is insufficient to reverse the magnetic moments the time of a measurement.
Here the susceptibility is the static susceptibility $\chi_1$. Between these two extremes there exists a maximum at the temperature $T_{\max}$. $\chi^\prime$ can be calculated from (\ref{Susc4}) using the formula for the relaxation time $\tau $ appropriate to the anisotropy symmetry, and considering a particular volume $V,$ one can determine the temperature $T_{\max}$.
In an assembly of particles with a volume distribution, $\chi^\prime$ can be calculated by postulating that at a given temperature and given measuring time, certain particles are in the superparamagnetic state and that the others are in the blocked state. The susceptibility is then given by the sum of the two contributions
\begin{equation}\label{Susc5}
\chi ^{\prime }(T,\nu )=\int\limits_{V_c}^{\infty }\mathcal{D}V\,\chi_{1}(T,V,\nu ) + \int\limits_0^{V_c}\mathcal{D}V\,\chi_0(T,V,\nu ),
\end{equation}
where $V_c=V_c(T,H)$ is the ``critical volume" defined as the volume that discriminates between superparamagnetic particles of volume $V < V_c$ and blocked particles with $V >V_c$, and is experiment-dependent.
Equation (\ref{Susc5}) is then rewritten for the ZFC magnetization as follows
\begin{eqnarray}
M_{zfc}(H,T,\psi) &=& \int\limits_{0}^{V_{c}}\mathcal{D}V\,\,M_{sp}(H,T,V,\psi) \nonumber\\
&+&\int\limits_{V_{c}}^{\infty }\mathcal{D}V\,\,M_{b}(H,T,V,\psi)
\end{eqnarray}
where $M_{sp}$ and $M_b$ are the contributions to the magnetization from the superparamagnetic and blocked particles, respectively. $\psi$ is the angle between the applied field and the anisotropy easy axis of each particle.

The critical volume $V_c$ can also be defined as the volume at which the relaxation rate of the system is equal to the measuring frequency, i.e., $\Gamma = \nu_m$.
In Ref.~\onlinecite{azekac06prep} we compute the contribution of DDI to $M_{sp}$ and $M_b$ and use the relaxation rate in (\ref{DDIRR}) to compute the DDI contribution to $V_c$. Then, we compute $T_{\max}$ as a function of the applied field for increasing intensity of DDI, i.e., increasing DDI parameter $\xi_d$, within the limit of validity discussed above.

\begin{figure}[ht!]
\includegraphics[width=8cm]{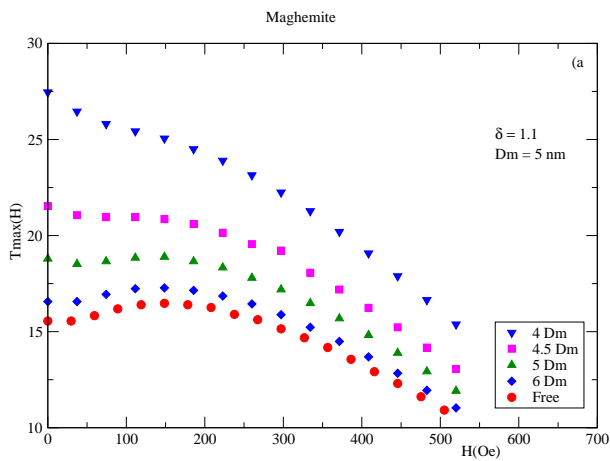}
\caption{The temperature $T_{\max}(H)$ as a function of the applied field for maghemite.}
\label{TmaxH_Maghemite5nm_delta1p1_DDI}
\end{figure}

In Fig.~\ref{TmaxH_Maghemite5nm_delta1p1_DDI} we present the results for an assembly of maghemite particles with mean diameter $D_m = 5$ nm and a distribution width $\delta=1.1$.
The curves are for a center-to-center inter-particle distance that is a multiple of $D_m$. The curve in closed circles is for the non-interacting (free) assembly.
%
%
We observe that indeed the effect of DDI is to change $T_{\max}(H)$ from a bell-like curve with a maximum to a monotonic decreasing function, and this compares well with the experimental results [see Fig. 1 of Ref.~\onlinecite{kacetal00jpcm}].
As was stressed in Ref.~\onlinecite{jongar01epl} the effect of DDI is not only to change the potential energyscape, as was argued in many previous publications \cite{doretal88jpc, mortro94prl, doretal97acp}, but also to introduce a transverse field that induces saddle points in the potential \cite{garetal99pre}.
This in turn makes the relaxation rate quite sensitive to the damping strength, and for low damping, as is the case in Fig.~\ref{TmaxH_Maghemite5nm_delta1p1_DDI} ($\lambda=0.1$), the probability of switching increases.
Then, if a magnetic field is added with increasing intensity the energy barrier is lowered and the magnetic moments switch at lower temperatures.
\subsection{Assembly of MSP particles}
In this last section we briefly comment on some recent (raw) results of the Monte Carlo calculations of the magnetization of a polydisperse assembly of MSP nanoparticles as dealt with in section \ref{Intrinsic}.
So far we have only considered non-interacting assemblies, and the work on the effect of DDI is still in progress.
%
\begin{figure}[floatfix]
\includegraphics[width=6cm, angle=-90]{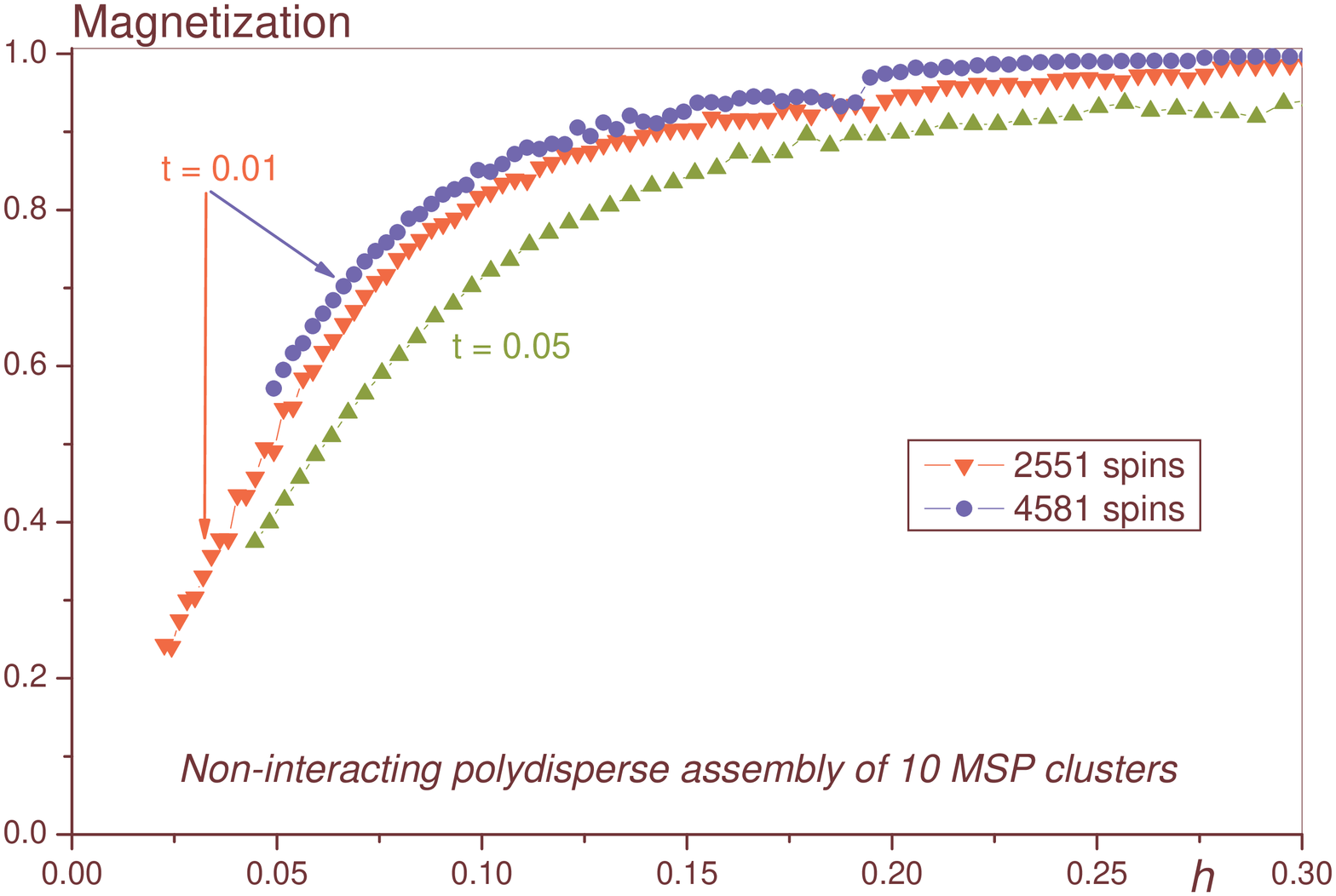}
\caption{Magnetization of a polydisperse assembly of MSP particles with uniaxial anisotropy in the core ($k_c=K_c/J=0.0024$) and TSA ($k_s=K_s/J=0.4$). $h=\mu_B H/J,\,t=k_BT/J$. The curve in up triangles is for ${\cal N}=2551$ and $t=0.05$.}
\label{mh_PolyMSP_2551_4581_t001_005}
\end{figure}
%
In Fig.~\ref{mh_PolyMSP_2551_4581_t001_005} we plot the (reduced) magnetization of two assemblies of $10$ particles each. The total number of atomic spins in these assemblies is $2551$ and $4581$. As the number of particles is the same in both assemblies, the particles in the assembly of $2551$ spins are smaller.
These preliminary results show that the magnetization of the assembly with smaller particles requires higher fields to saturate even at very low temperature, as can be seen by comparing the two upper curves for $t=0.01$ which corresponds to $T=1$ K.
As the temperature increases the spin disorder caused by the boundary and surface effects becomes stronger leading to a decrease of the magnetization [see the curve at $t=0.05$.]
\section{Conclusions}
Contrary to the introduction, the conclusion is kept rather short. We will limit ourselves to emphasizing the unresolved issues owing to the complexities inherent to the systems under study, to the lack of sufficient experimental information, and to the deficiencies of the actual models.

Nonetheless, we may confort ourselves saying that so far within the framework of the present models we have achieved a fair understanding of the static properties of nanomagnets, namely the spatial distribution of the magnetization and its switching mechanisms. In particular, it becomes clearer, to both experimentalists and theorists, that the many-spin aspect of the nanomagnets is a prerequisite for a better understanding of their properties when their size is reduced towards the nanometer.
We believe that the work done so far by many authors has set the pace to future faster developments towards mastering and controlling the novel extraordinary features of nanomagnets in view of efficient and practical technological applications. From the viewpoint of fundamental physics the facts established so far constitute a stepstone towards ever richer areas with more  challenges to come.

However, a great deal of work and endeavor still lies ahead of us because our actual knowledge of the subject is still plagued with quite a few obscure areas owing to the lack of sufficient pertinent information. This of course prevents us today from building a complete picture of the puzzle and makes our approach rather approximate.
Information is gained by experimental investigation along with theoretical predictions and attempts of interpretation.
At present, the theoretical models developed so far present a few deficiencies.
For instance, it is commonly assumed that the crystal structure on the surface is the same as in the core with the same atomic lattice parameters. This cannot be wholly true considering the possibility of surface reconstructions. Of course, some models, including those presented here, do include apices, edges and facets, and the possibility of taking the exchange coupling on the surface as different from that in the core, or that between the core and the surface.
However, there is no guarantee that it will become experimentally possible in the near future to estimate the atomic positions and lattice parameters and may be theoretically possible to perform crystal field calculations, and eventually check these assumptions.
The magneto-crystalline anisotropy constant and exchange coupling in the core are also taken as those in the bulk underlying material. However, we have shown that the core of a nanoparticle does not enjoy the properties of the underlying bulk material.
For a given material such parameters could vary with the radius of the particle. In addition, the intensity and nature of surface anisotropy constitute a real challenge. There are many estimates but no firm understanding is achieved yet as to how surface anisotropy stems from the atomic structure in nanoparticles of reasonable size.

For these interrogations to receive answers experimentalists will have to devise ever more sensitive equipments and measurement techniques in order to probe the intrinsic properties of nanomagnets and have direct access to the related physical observables. This, of course, will require new strategies for further isolating the nanomagnets and rid them off the influence of their hosting matrices and mutual interactions.
On the other hand, understanding the influence on the particles static and dynamic behavior of the surrounding matrix and the inter-particle interactions, is of paramount importance to efficient practical applications.
%
%

%
\end{document}